\documentclass[10pt,journal,compsoc]{IEEEtran}

\usepackage{graphicx}
\usepackage{algorithm}
\usepackage[noend]{algpseudocode}
\usepackage{mathtools}
\usepackage{amsfonts}

\usepackage[bookmarks=true,breaklinks=true,letterpaper=true,colorlinks,linkcolor=black,citecolor=blue,urlcolor=black]{hyperref}

\begin{document}

\title{\huge Optimizing Routerless Network-on-Chip Designs:\linebreak An Innovative Learning-Based Framework}

\author{\IEEEauthorblockN{
Ting-Ru Lin\IEEEauthorrefmark{2},
Drew Penney\IEEEauthorrefmark{1},
Massoud Pedram\IEEEauthorrefmark{2},
Lizhong Chen\IEEEauthorrefmark{1}$^{,}$\IEEEauthorrefmark{4}
}

\thanks{\IEEEauthorrefmark{4}Corresponding author. Email: chenliz@oregonstate.edu}
\thanks{\IEEEauthorrefmark{2}The authors are with University of Southern California, CA 90007}
\thanks{\IEEEauthorrefmark{1}The authors are with Oregon State University, OR 97331}

}

\IEEEtitleabstractindextext{%

\begin{abstract}
Machine learning applied to architecture design presents a promising opportunity with broad applications. Recent deep reinforcement learning (DRL) techniques, in particular, enable efficient exploration in vast design spaces where conventional design strategies may be inadequate. This paper proposes a novel deep reinforcement framework, taking routerless networks-on-chip (NoC) as an evaluation case study. The new framework successfully resolves problems with prior design approaches being either unreliable due to random searches or inflexible due to severe design space restrictions. The framework learns (near-)optimal loop placement for routerless NoCs with various design constraints. A deep neural network is developed using parallel threads that efficiently explore the immense routerless NoC design space with a Monte Carlo search tree. Experimental results show that, compared with conventional mesh, the proposed deep reinforcement learning (DRL) routerless design achieves a 3.25x increase in throughput, 1.6x reduction in packet latency, and 5x reduction in power. Compared with the state-of-the-art routerless NoC, DRL achieves a 1.47x increase in throughput, 1.18x reduction in packet latency, and 1.14x reduction in average hop count albeit with slightly more power overhead.
\end{abstract}

}

\maketitle

\IEEEdisplaynontitleabstractindextext

\IEEEraisesectionheading{\section{Introduction}\label{sec:introduction}}
   Improvements in computational capabilities are increasingly reliant upon advancements in many-core chip designs. These designs emphasize parallel resource scaling and consequently introduce many considerations beyond those in single core processors. As a result, traditional design strategies may not scale efficiently with this increasing parallelism. 
   Early machine learning approaches, such as simple regression and neural networks, have been proposed as an alternative design strategy. More recent machine learning developments leverage deep reinforcement learning to provide improved design space exploration. This capability is particularly promising in broad design spaces, such as network-on-chip (NoC) designs.
   
   NoCs provide a basis for communication in many-core chips that is vital for system performance \cite{PacketsNotWires}. NoC designs involve many trade-offs between latency, throughput, wiring resources, and other overhead. Exhaustive design space exploration, however, is often infeasible in NoCs and architecture design in general due to immense design spaces. Thus, intelligent exploration approaches would greatly improve and benefit NoC designs.
   
   Applications include recently proposed novel routerless NoCs \cite{IMR, RL}. Conventional router-based NoCs incur significant power and area overhead due to complex router structures. Routerless NoCs eliminate these costly routers by effectively using wiring resources while achieving comparable scaling to router-based NoCs. Prior research has demonstrated up to 9.5x reduction in power and 7x reduction in area compared with Mesh \cite{RL}, establishing routerless NoCs as a promising alternative for NoC designs. Like many novel concepts and approaches in the architectural field, substantial ongoing research is needed to explore the full potential of the routerless NoC design paradigm. Design challenges for routerless NoCs include efficiently exploring the huge design space (easily exceeding $10^{12}$) while ensuring connectivity and wiring resource constraints. This makes routerless NoCs an ideal case study for intelligent design exploration approaches.

   Routerless NoC approach has, thus far, followed two approaches. The first, Isolated Multi-Ring (IMR) \cite{IMR}, uses an evolutionary approach (genetic algorithm) for loop design based on random mutation/exploration. The second approach (REC) \cite{RL} recursively adds loops following a strict methodology based on the NoC size, thus severely restricting broad applicability. Briefly, neither approach guarantees efficient generation of fully-connected routerless NoC designs under various constraints.

  In this paper, we propose a novel deep reinforcement learning framework for design space exploration, and demonstrate a specific implementation using routerless NoC design as our case study. Efficient design space exploration is realized using a Monte-Carlo Tree Search (MCTS) that generates training data to a deep neural network which, in turn, guides the search in MCTS. Together, the framework self-learns loop placement strategies obeying design constraints. Evaluation shows that the proposed deep reinforcement learning design (DRL) achieves a 3.25x increase in throughput, 1.6x reduction in packet latency, and 5x reduction in power compared with a conventional mesh. Compared with REC, the state-of-the-art routerless NoC, DRL achieves a 1.47x increase in throughput, 1.18x reduction in packet latency, and 1.14x reduction in average hop count albeit with slightly more power overhead. When scaling from a 4x4 to a 10x10 NoC under synthetic workloads, the throughput drop is also reduced dramatically from 31.6\% in REC to only 4.7\% in DRL.
    
  Key contributions of this paper include: 
 
\begin{itemize}
\item Fundamental issues are identified in applying deep reinforcement learning to routerless NoC designs;
\item An innovative deep reinforcement learning framework is proposed and implementation is presented for routerless NoC design with various design constraints;
\item  Cycle-accurate architecture-level simulations and circuit-level implementation are conducted to evaluate the design in detail;
\item Broad applicability of the proposed framework with several possible examples is discussed.
\end{itemize}
  
  The rest of the paper is organized as follows: Section 2 provides background on NoC architecture and reinforcement learning techniques; Section 3 describes the issues in prior methods for routerless NoC problems and the need for a better method; Section 4 details the proposed deep reinforcement learning framework; Section 5 illustrates our evaluation methodology; Section 6 provides simulation results; Section 7 reviews related work; and Section 8 conclude the paper.

\section{Background}
\subsection{NoC Architecture}
\begin{figure}[ht]
\centering
\includegraphics[width=0.45\textwidth]{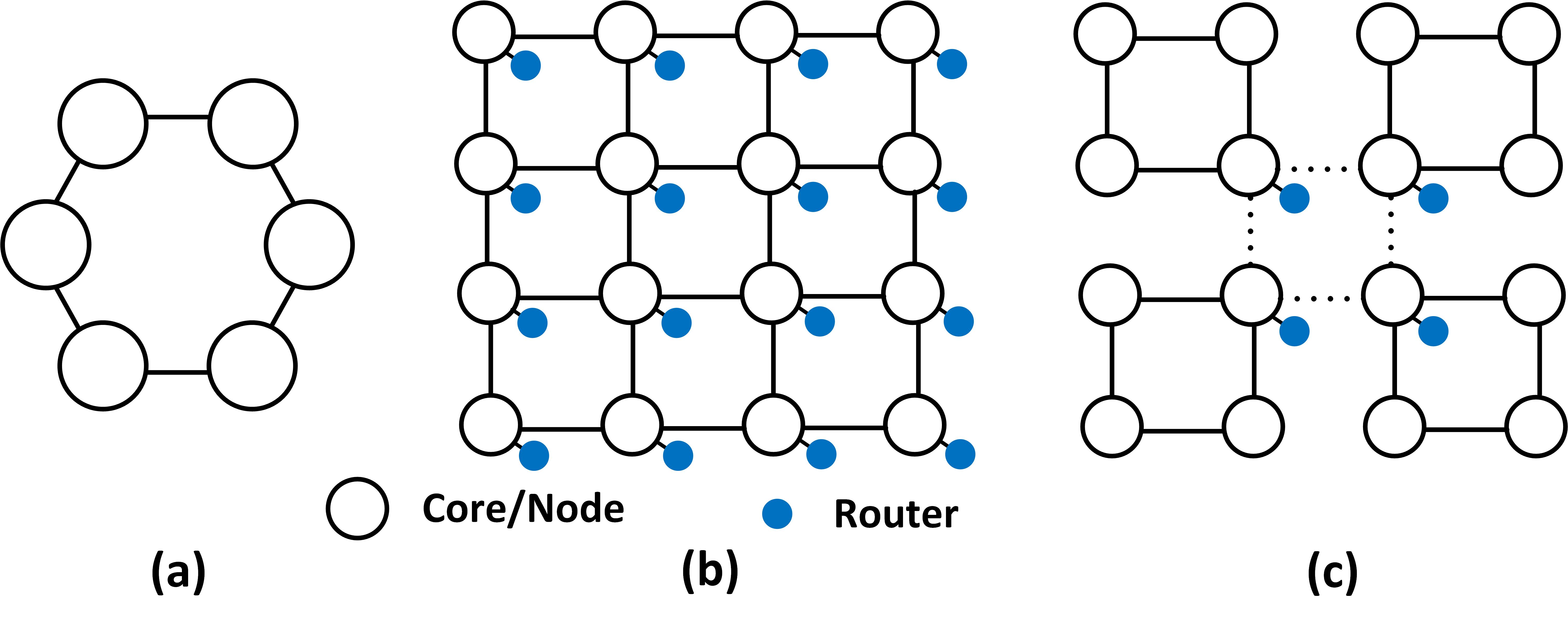}
\caption{NoC Architecture. (a) Single-Ring (b) Mesh (c) Hierarchical Ring}
\label{fig:NoCs}
\end{figure}

\textbf{Single-ring NoCs:} Nodes in a single-ring NoC communicate using one ring connecting all nodes.\footnote{Note that rings and loops are used interchangeably in this paper.} Packets are injected at a source node and forwarded along the ring to a destination node. An example single-ring NoC is seen in Figure \ref{fig:NoCs}(a). Single-ring designs are simple, but have low bandwidth capabilities, severely restricting their applicability in large-scale designs. Specifically, network saturation is rapidly reached as more nodes are added due to frequent end-to-end control packets \cite{William2007}. Consequently, most single-ring designs only scale to a modest number of processors \cite{NoCbook}. 

\textbf{Router-based NoCs:} Routers in NoC designs generally consist of input buffers, routing and arbitration logic, and a crossbar connecting input buffers output links. These routers enable a decentralized communication system in which routers check resource availability before packets are sent between nodes \cite{RL}. Mesh (or mesh-based architectures) are a common router-based NoC and have become the \textit{de facto} choice due to their scalability and relatively high bandwidth \cite{IMR}. The basic design, shown in Figure \ref{fig:NoCs}(b), features a 2D grid of nodes with a router at every node.
These routers can incur 11\% chip area \cite{Gratz2006} and, depending upon frequency and activity, up to 28\% chip power \cite{NoRD2012,Mesh} overhead (although some recent work \cite{PITON, PITON2016} has shown a much smaller router overhead when configured with narrow links and shallow/few buffers at the cost higher latency; this indirectly shows that routers are the main cost in existing NoCs). Hierarchical-ring is a common multi-ring design that uses a hierarchy of local and global rings. Figure \ref{fig:NoCs}(c) illustrates this hierarchy in which the dotted global ring connects as local rings together. Routers are only needed for nodes intersected by the global ring as they are responsible for packet transfer between ring groups \cite{Rachata2014}. Extensive research has explored router-based NoC optimization \cite{NoRD2012,Udipi2010,Howard2011}, but these solutions can only slightly reduce power and area overhead \cite{IMR}.

\textbf{Routerless NoCs:} Significant overhead associated with router-based topologies has motivated routerless NoC designs. Early proposals \cite{Udipi2010} used bus-based networks in a hierarchical approach by dividing the chip into multiple segments, each with a local broadcast bus. Segments are connected by a central bus with low-cost switching elements. These bus-based networks inevitably experience contention on local buses and at connections with the central bus, resulting in poor performance under heavy traffic. Recently, isolated multi-ring (IMR) NoCs have been proposed that exploit additional interconnect wiring resources available in modern semiconductor processes \cite{IMR}. Nodes are connected via at least one ring and packets are forwarded from source to destination without switching rings. IMR improves over mesh-based designs in terms of power, area, and latency, but requires significant buffer resources: each node has a dedicated input buffer for each ring passing through its interface, thus a single node may require many packet-sized buffers \cite{IMR,RL}. Recent routerless NoC design (REC) \cite{RL} has mostly eliminated these costly buffers by adopting shared packet-size buffers among loops. REC uses just a single flit-sized buffer for each loop, along with several shared extension buffers to provide effectively the same functionality as dedicated buffers \cite{RL}. 
  
\begin{figure}[ht]
\centering
\includegraphics[width=0.47\textwidth]{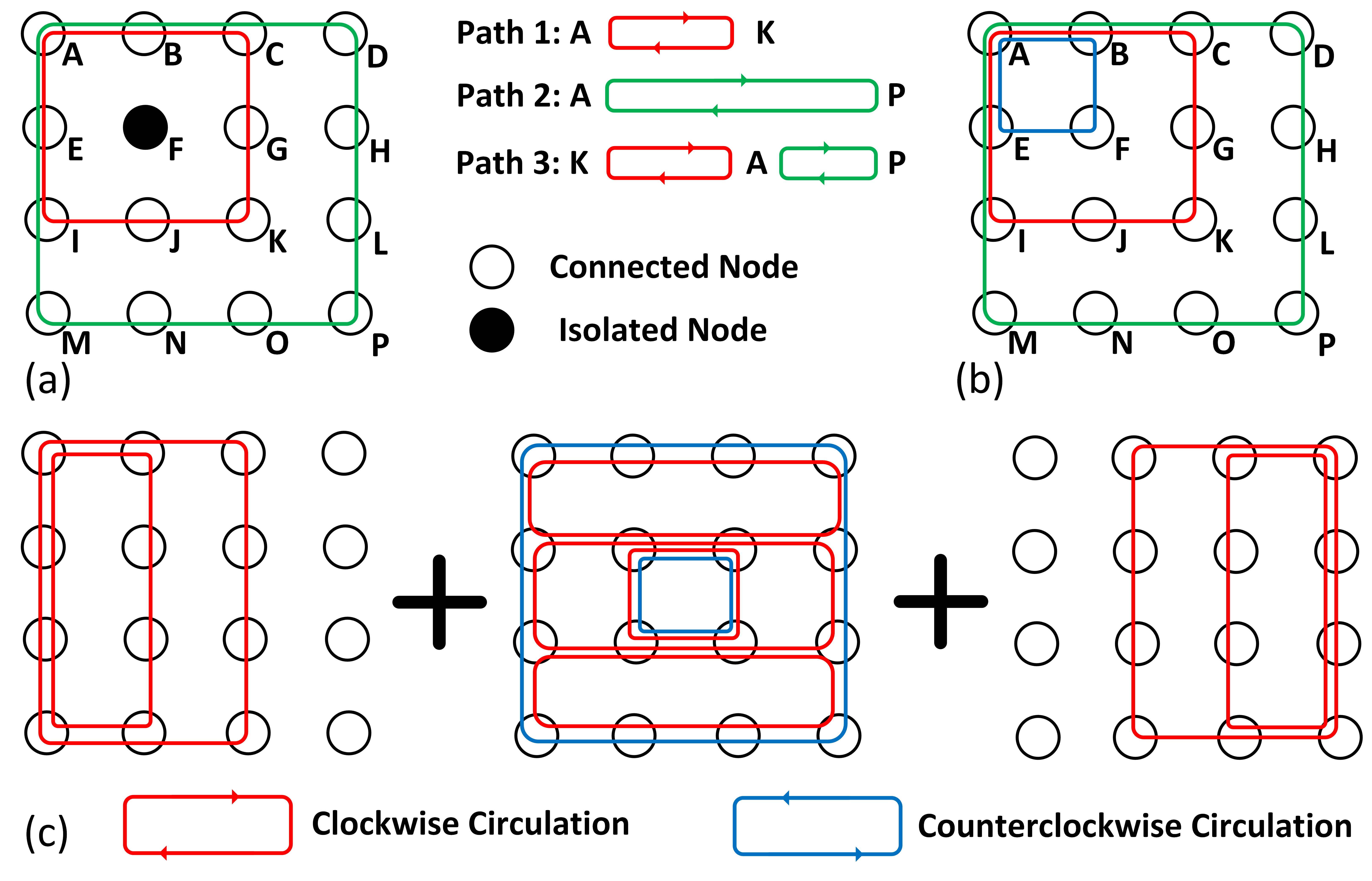}
\caption{A 4x4 NoC with rings. (a) A NoC with one isolated node. (b) A NoC without isolated nodes. (c) A 4x4 routerless NoC with rings.} 
\label{fig:MRNoC}
\end{figure}

Both IMR and REC routerless NoC designs differ from previous designs in that no routing is performed during traversal, so packets in a loop cannot be forwarded to another loop \cite{IMR,RL}. Both designs must therefore satisfy two requirements: every pair of nodes must be connected by at least one loop and all routing is done at the source node. Figure \ref{fig:MRNoC} delineates these requirements and highlights differences between router-based and routerless NoC designs. Figure \ref{fig:MRNoC}(a) depicts an incomplete 4x4 ring-based NoC with three loops. These loops are unidirectional so arrows indicate the direction of packet transfer for each ring. Node $F$ is isolated and cannot communicate with other nodes since no ring passes through its interface. Figure \ref{fig:MRNoC}(b) depicts the NoC with an additional loop through node $F$. If routers are used, such as at node $A$, this ring would complete the NoC, as all nodes can communicate with ring switching. Packets from node $K$, for example, can be transferred to node $P$ using path 3, which combines $paths 1$ and $path 2$. In a routerless design, however, there are still many nodes that cannot communicate as packets must travel along a \textit{single} ring from source to destination. That is, packets from node $K$ cannot communicate with node $P$ because $path 1$ and $path 2$ are isolated from each other. Figure \ref{fig:MRNoC}(c) depicts an example REC routerless NoC for 4x4\cite{RL}. Loop placement for larger networks is increasingly challenging.

  Routerless NoCs can be built with simple hardware interfaces by eliminating crossbars and VC allocation logic. As a result, current state-of-the-art routerless NoCs have achieved 9.5x power reduction, 7.2x area reduction, and 2.5x reduction in zero-load packet latency compared with conventional mesh topologies \cite{RL}. Packet latency, in particular, is greatly improved by single-cycle delays per hop, compared to standard mesh, which usually requires two cycles for the router alone. Hop count in routerless designs can asymptotically approach the optimal mesh hop count using additional loops at the cost of power and area. Wiring resources, however, are finite, meaning that one must restrict the total number of overlapping rings at each node (referred to as node overlapping) to maintain physical realizability. In Figure \ref{fig:MRNoC} (b), node overlapping at node $A$, for example, is three, whereas node overlapping at node $F$ is one. Routerless NoC design is a trivial task if wiring resources are ignored, but, with strict node overlapping, design becomes substantially more challenging. As discussed in Section 3, existing methods either do not satisfy or do not enforce these potential constraints. We therefore explore potential applications of machine learning to design constraints such as node overlapping.
  
\subsection{Reinforcement Learning}
\begin{figure}[ht]
\centering
\includegraphics[width=0.43\textwidth]{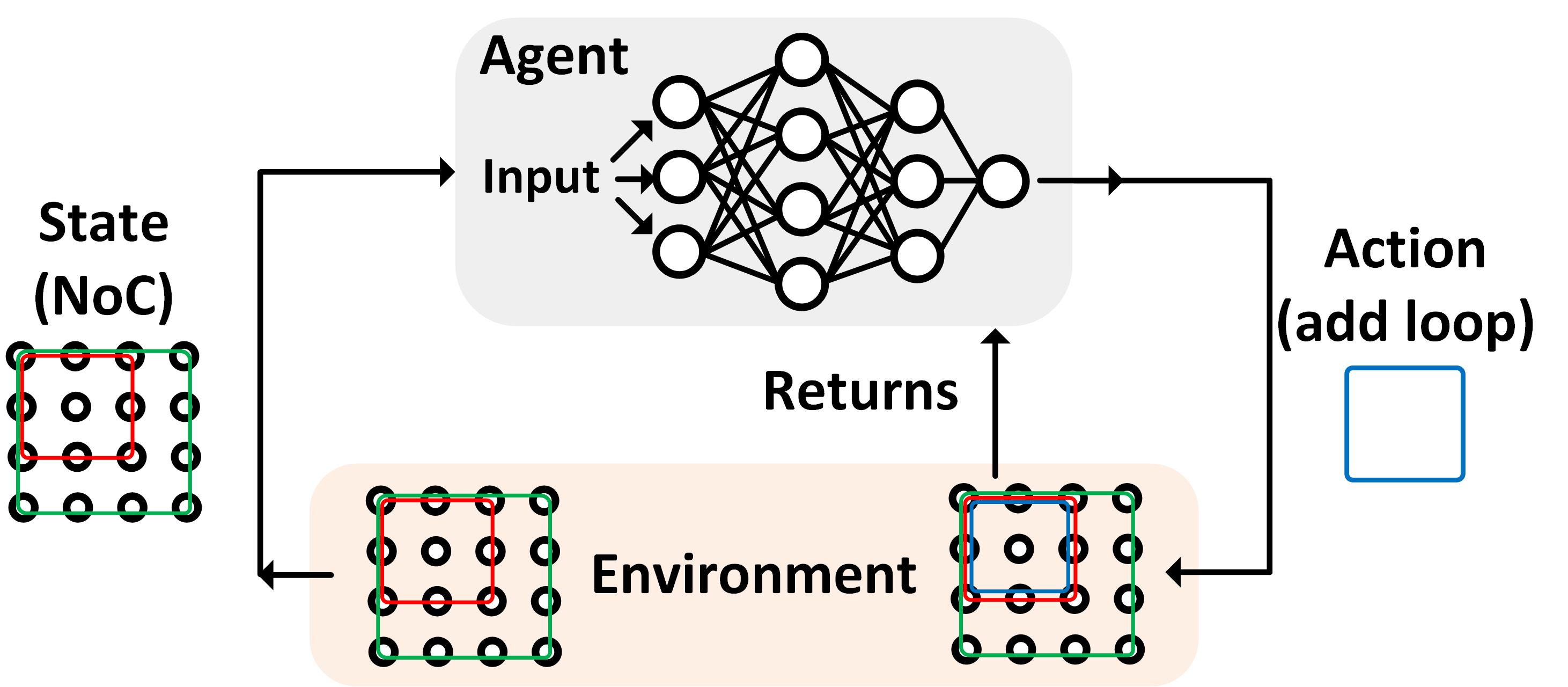}
\caption{Reinforcement learning framework.}
\label{fig:RLframework}
\end{figure}

\textbf{Background of Reinforcement Learning:} 
Reinforcement learning is a branch of machine learning that explores actions in an environment to maximize cumulative returns/rewards. Fundamental to this exploration is the environment, $\mathcal{E}$, in which a software agent takes actions. In our paper, this environment is represented by a routerless NoC design. The agent attempts to learn an optimal policy $\pi$ for taking a sequence of actions $\{a_t\}$ from each state $\{s_t\}$, acquiring returns $\{r_t\}$ at different times $t$ in $\mathcal{E}$ \cite{RLbook}. Figure \ref{fig:RLframework} depicts the exploration process in which the agent learns to take an action $a_t$ (adding a loop) given a state $s_t$ (information about an incomplete routerless NoC) with the end goal of maximizing returns (minimizing average hop count). The agent is encouraged to explore a broad set of states in order to maximize cumulative returns. At each of these states, there is a transition probability, $P(s_{t+1}|s_t, a_t)$, which represents the probability of transitioning from $s_t$ to $s_{t+1}$ given $a_t$. The learned value function $V^\pi(s)$ under policy $\pi$ is represented by 
\begin{align}
\label{eq:optV}
  & V^\pi({s})= \mathbb{E} [\sum_{t\geq0}\gamma^t*r_t | s_0=s, \pi]\\
\label{eq:optV2}
  & R = \sum_{t\geq0}\gamma^t*r_t
\end{align}
where $\gamma$ is a discount factor ($\leq 1$) and $R$ is the discounted cumulative return. 

  The goal of reinforcement learning is to maximize cumulative returns $R$ and, in case of routerless NoC design, to minimize average hop count. To this end, the agent attempts to learn the optimal policy $\pi^*$ that satisfies 
\begin{equation}
\label{eq:OptPolicy}
  \pi^*({s})= arg \max_\pi\mathbb{E} [\sum_{t\geq0}\gamma^t*r_t |  s_0=s, \pi].
\end{equation} 
Equation \ref{eq:optV} under $\pi^*$ thus satisfies the Bellman equation
\begin{align}
\label{eq:OptPolicy1}
  & V^*({s})= \mathbb{E} [r_0 + \gamma V^*(s_1) | s_0=s, \pi^*]\\
\label{eq:OptPolicy2}
  &\quad = p(s_0)\sum_{a_0}\pi^*(a_0; s_0)\sum_{s_1}P(s_1|s_0, a_0)[r(s_0, a_0)+\gamma V^*(s_1)],
\end{align}
where $p(s_0)$ is the probability of an initial state $s_0$. Equation \ref{eq:OptPolicy2} suggests that an agent, after learning the optimal policy function $\pi^*$, can minimize the average hop count of a routerless NoC. The set of possible loops, however, poses a significant challenge for reinforcement learning exploration. For instance, there are over a trillion ($\binom{784}{5}=2.44 \times 10^{12})$ ways to choose just five loops from the 784 possible rectangular loops for an 8x8 NoC. This challenge requires efficient exploration and a powerful approximator that can learn an appropriate policy function and/or value function.

\textbf{Deep Reinforcement Learning:} 
  Breakthroughs in deep learning have spurred researchers to rethink potential applications for deep neural networks (DNNs) in diverse domains. One result is deep reinforcement learning, which synthesizes DNNs and reinforcement learning concepts to address complex problems \cite{Mnih2013,Silver2016,Silver2017}. This synthesis mitigates data reliance without introducing convergence problems via efficient data-driven exploration based on DNN output. Recently, these concepts have been applied to Go, a grid-based strategy game involving stone placement. In this model, a trained policy DNN learns optimal actions by searching a Monte Carlo tree that records actions suggested by the DNN during training \cite{Silver2016,Silver2017}. Deep reinforcement learning can outperform typical reinforcement learning by generating a sequence of actions with better cumulative returns \cite{Mnih2013,Silver2016,Silver2017}. 
  
\section{Motivation}  
\subsection{Design Space Exploration}
  Deep reinforcement learning provides a powerful foundation for design space exploration using continuously refined domain knowledge. This capability is advantageous since existing methods for routerless NoC designs have limited design space exploration capabilities. Specifically, the evolutionary approach \cite{IMR} explores the design space by evaluating generations of individuals and offspring. Selection uses an objective function while evolution relies on random mutation, leading to an unreliable search since past experiences are ignored. Consequently, exploration can be misled and, when applied to routerless NoCs, generate configurations with high average hop count and long 48-hop loops in an 8x8 NoC \cite{RL}. The recursive layering approach (REC) overcomes these reliability problems using a predefined configuration for each network size, thus strictly limiting design flexibility. Latency improves as the generated loops pass through fewer nodes on average\cite{RL}, but hop count still suffers in comparison to router-based NoCs as it is restricted by the total number of loops. For an 8x8 NoC, the average hop count is 5.33 in mesh and 8.32 in the state-of-the-art recursive layering design, corresponding to a 1.5x increase \cite{RL}.
  
  Both approaches are also limited by their inability to enforce design constraints, such as node overlapping. Specifically, node overlapping is a byproduct of their algorithms, rather than a design constraint. In IMR, ring selection is based solely on inter-core-distance and ring lengths \cite{IMR} so node overlapping may vary significantly based on random ring mutation. Constraints could be built into the fitness function, but these constraints are likely to be violated to achieve better performance. Alternatively, in REC, loop configuration for each network size is strictly defined. A 4x4 NoC must use exactly the loop structure shown in Figure \ref{fig:MRNoC} (c) so node overlapping cannot be changed without modifying the algorithm itself. These constraints must be considered during loop placement since an optimal design will approach these constraints to allow many paths for packet transfer. 
  
\subsection{Reinforcement Learning Challenges}
  Several considerations apply to deep reinforcement learning in any domain. To be more concrete, we discuss these considerations in the context of routerless NoC designs.
  
\textbf{Specification of States and Action:} State specification must include all information for the agent to determine optimal loop placement and should be compatible with DNN input/output structure. An agent that attempts to minimize average hop count, for example, needs information about the current hop count. Additionally, information quality can impact learning efficiency since inadequate information may require additional inference. Both state representation and action specification should be a constant size throughout the design process because the DNN structure is invariable. 
  
\textbf{Quantification of Returns:} Return values heavily influence NoC performance so they need to encourage beneficial actions and discourage undesired actions. For example, returns favoring large loops will likely generate a NoC with large loops. Routerless NoCs, however, benefit from diverse loop sizes; large loops help ensure high connectivity while smaller loops may lower hop counts. It is difficult to achieve this balance since the NoC will remain incomplete (not fully connected) after most actions. Furthermore, an agent may violate design constraints if the return values do not appropriately deter these actions. Returns should be conservative to discourage useless or illegal loop additions.

\textbf{Functions for Learning:} Optimal loop configuration strategies are approximated by learned functions, but these functions are notoriously difficult to learn due to high data requirements. This phenomenon is observed in AlphaGo \cite{Silver2016} where the policy function successfully chooses from $19^2$ possible moves at each of several hundred steps, but requires more than 30 million data samples. An effective approach must consider this difficulty, which can be potentially addressed with optimized data efficiency and parallelization across threads, as discussed later in our approach.

\textbf{Guided Design Space Search:} An ideal routerless NoC would maximize performance while minimizing loop count based on constraints. Similar hop count improvement can be achieved using either several loops or a single loop. Intuitively, the single loop is preferred to reduce NoC resources, especially under strict overlapping constraints. This implies benefits from ignoring/trimming exploration branches that add loops with suboptimal performance improvement.

\section{Proposed Scheme}
\subsection{Overview}
\begin{figure}[ht]
\centering
\includegraphics[width=0.48\textwidth]{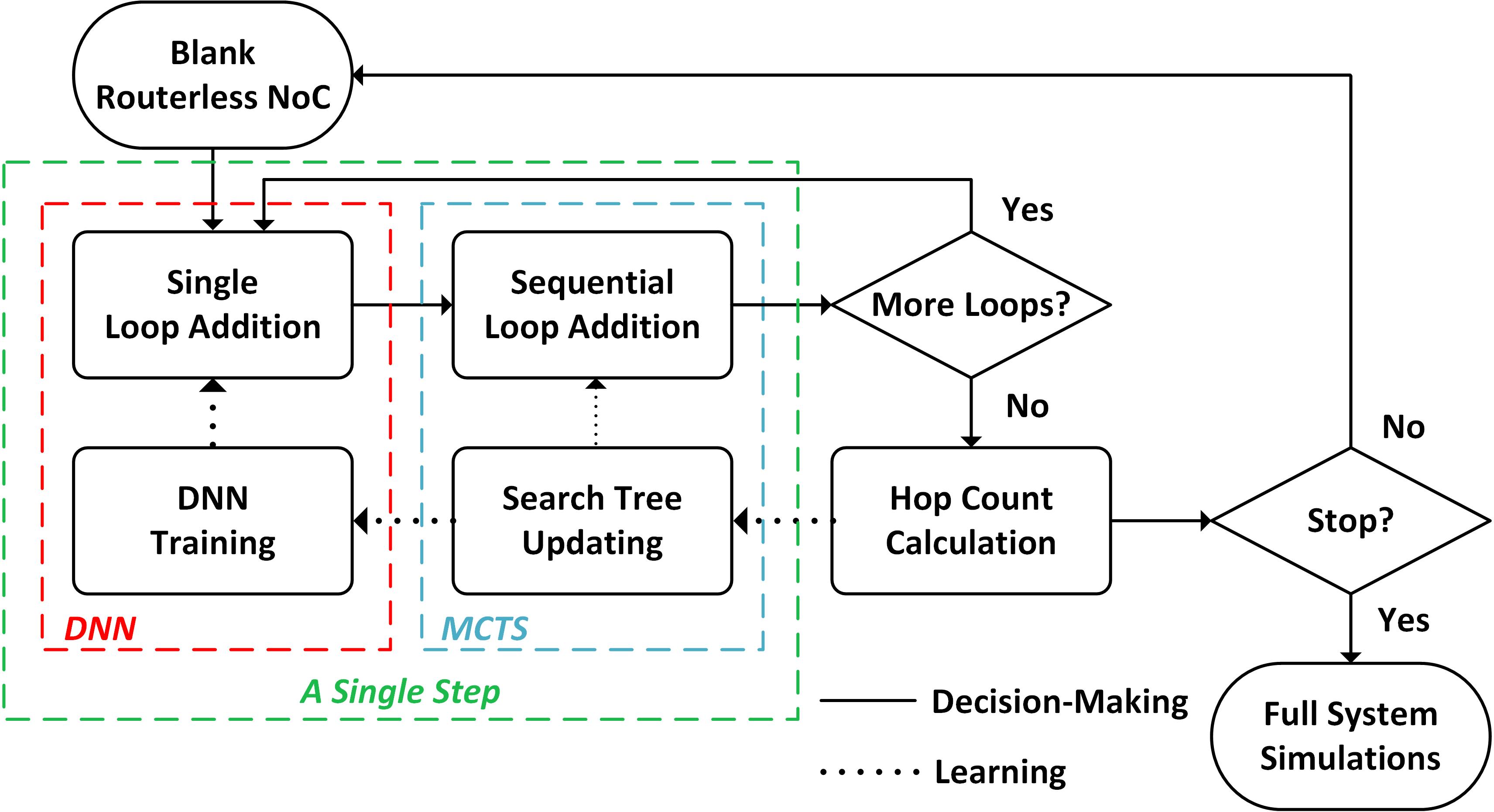}
\caption{Proposed deep reinforcement learning framework for routerless NoC designs.}
\label{fig:framwork}
\end{figure}

  The proposed deep reinforcement learning framework is depicted in Figure \ref{fig:framwork}. Framework execution begins by initializing the Monte Carlo Tree Search (MCTS) with an empty tree and a neural network without $a$ $priori$ training. The whole process consists of many exploration cycles. Each cycle begins with a completely disconnected NoC and adds loops (actions) in steps until the NoC is fully connected. As shown in the figure, several loops may be added in one step. The DNN (dashed "DNN" box) selects a good initial loop, which in a sense directs the search to a particular region in the design space; then several additional loops are added by following MCTS (dashed "MCTS" box) in that region. The MCTS starts from the current NoC layout (a MCTS node), and tree traversal selects loop placements using either greedy exploration with a probability $\epsilon$ or an "optimal" action until a leaf (one of many explored NoC configurations) is reached. Additional steps can be taken to add more loops and reach a fully connected network. At the end of the cycle, an overall reward (e.g., based on hop count) is calculated and combined with information on state, action, and value estimates to train the neural network and update the search tree (the dotted "Learning" lines). The exploration cycle repeats many times until a (near-)optimal routerless NoC design is found ("Stop"). Once the search completes, full system simulations are used to verify and evaluate the design. In the framework, the DNN generates coarse designs while MCTS efficiently refines these designs based on prior knowledge to continuously improve NoC configurations. Different from traditional supervised learning, the framework does not require a training dataset beforehand; instead, the DNN and MCTS are gradually trained by themselves from past exploration cycles.
  
  The above actions, rewards, and state representations in the proposed framework can be generalized for design space exploration in router-based NoCs and in other NoC-related research. Several generalized framework examples are discussed in Section \ref{BroadApp}. The remainder of this section addresses the application of the framework to routerless NoC design as a way to present low-level design and implementation details. Other routerless NoC implementation details including deadlock, livelock, and starvation are addressed in previous work \cite{IMR, RL} so are omitted here.

\subsection{Routerless NoCs Representation} \label{representation}
\begin{figure}[ht]
\centering
\includegraphics[width=0.4\textwidth]{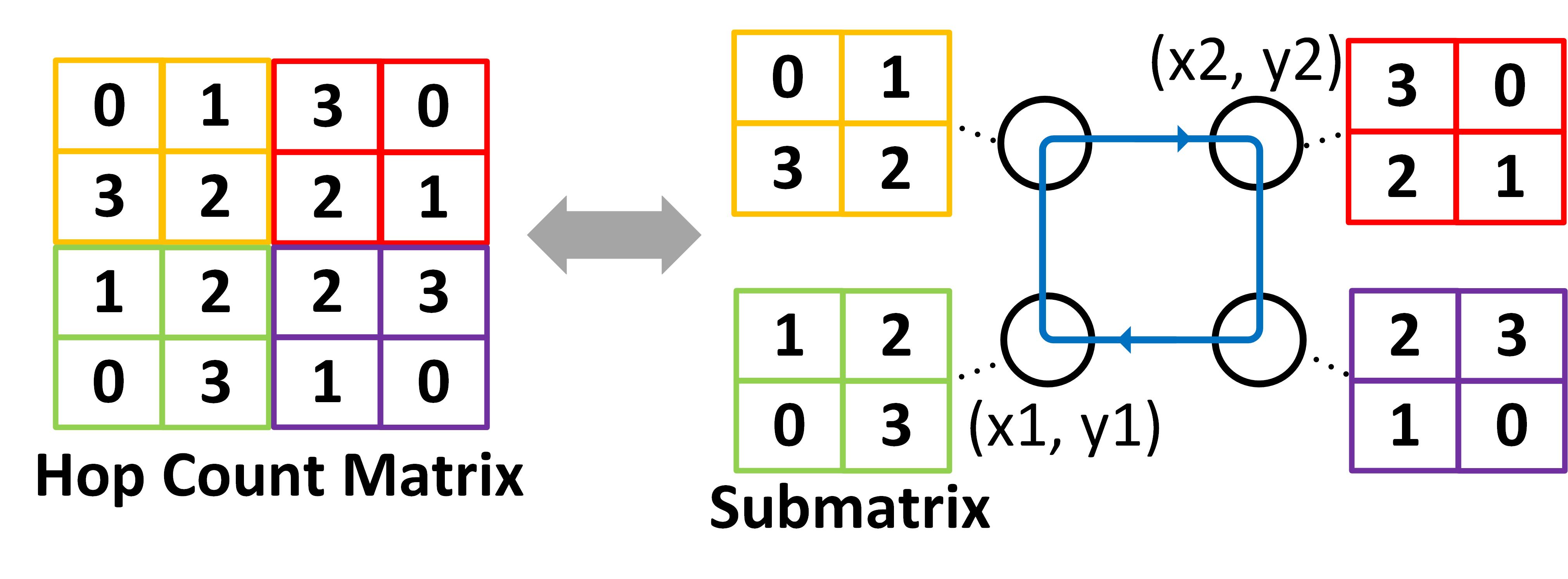}
\caption{Hop count matrix of a 2x2 routerless NoC.}
\label{fig:Hopcount}
\end{figure}

\textbf{Representation of Routerless NoCs (States):} State representation in our framework uses a hop count matrix to encode current NoC state as shown in Figure \ref{fig:Hopcount}. A 2x2 routerless NoC with a single clockwise loop is considered for simplicity. The overall state representation is a $4x4$ matrix composed of four 2x2 submatrices, each representing hop count from a specific node to every node in the network. For example, in the upper left submatrix, the zero in the upper left square corresponds to distance from the node to itself. Moving clockwise with loop direction, the next node is one hop away, then two, and three hops for nodes further along the loop. All other submatrices are generated using the same procedure. This hop count matrix encodes current loop placement information using a fixed size representation to accommodate fixed DNN layer sizes. In general, the input state for an $N\,x\,N$ NoC is an $N^2\,x\,N^2$ hop count matrix. Connectivity is also implicitly represented in this hop count matrix by using a default value of $5*N$ for unconnected nodes.

\textbf{Representation of Loop Additions (Actions):} An action is defined as adding a loop to an $N \times N$ NoC. We choose to restrict loops to rectangles to minimize the longest path. With this restriction, the longest path will be between diagonal nodes at the corners of the NoC, as in REC \cite{RL}. Actions are encoded as $(x1, y1, x2, y2, dir)$ where $x1, y1, x2$ and $y2$ represent coordinates for diagonal nodes $(x1, y1)$ and $(x2, y2)$ in the NoC, and $dir$ indicates packet flow direction within a loop. $dir>0.5$ represents clockwise circulation for packets whereas $dir\leq 0.5$ represents counterclockwise circulation. For example, the loop in Figure \ref{fig:Hopcount} represents the action $(0, 0, 1, 1, 1)$. We enforce rectangular loops by checking that $x1\neq x2$ and $y1\neq y2$.

\subsection{Returns After Loop Addition} 
  The agent is encouraged to fully explore the state space using a reward function that returns zero for all valid actions. Conversely, the agent is discouraged from taking repetitive, invalid, or illegal actions using negative returns (penalties). A repetitive action refers to adding a duplicate loop, receiving a $-1$ penalty. An invalid action refers to adding a non-rectangular loop, corresponding to a $-1$ penalty. Finally, an illegal action refers to adding a loop that causes the node overlapping constraint of $2*(N-1)$ to be violated, resulting in a severe $-5*N$ penalty. The agent receives a final return to characterize overall performance by subtracting average hop count in the generated NoC from average mesh hop count. Minimal average hop count is therefore found by minimizing the magnitude of cumulative returns.
  
\subsection{Deep Neural Network}
\begin{figure}[ht]
\centering
\includegraphics[width=0.45\textwidth]{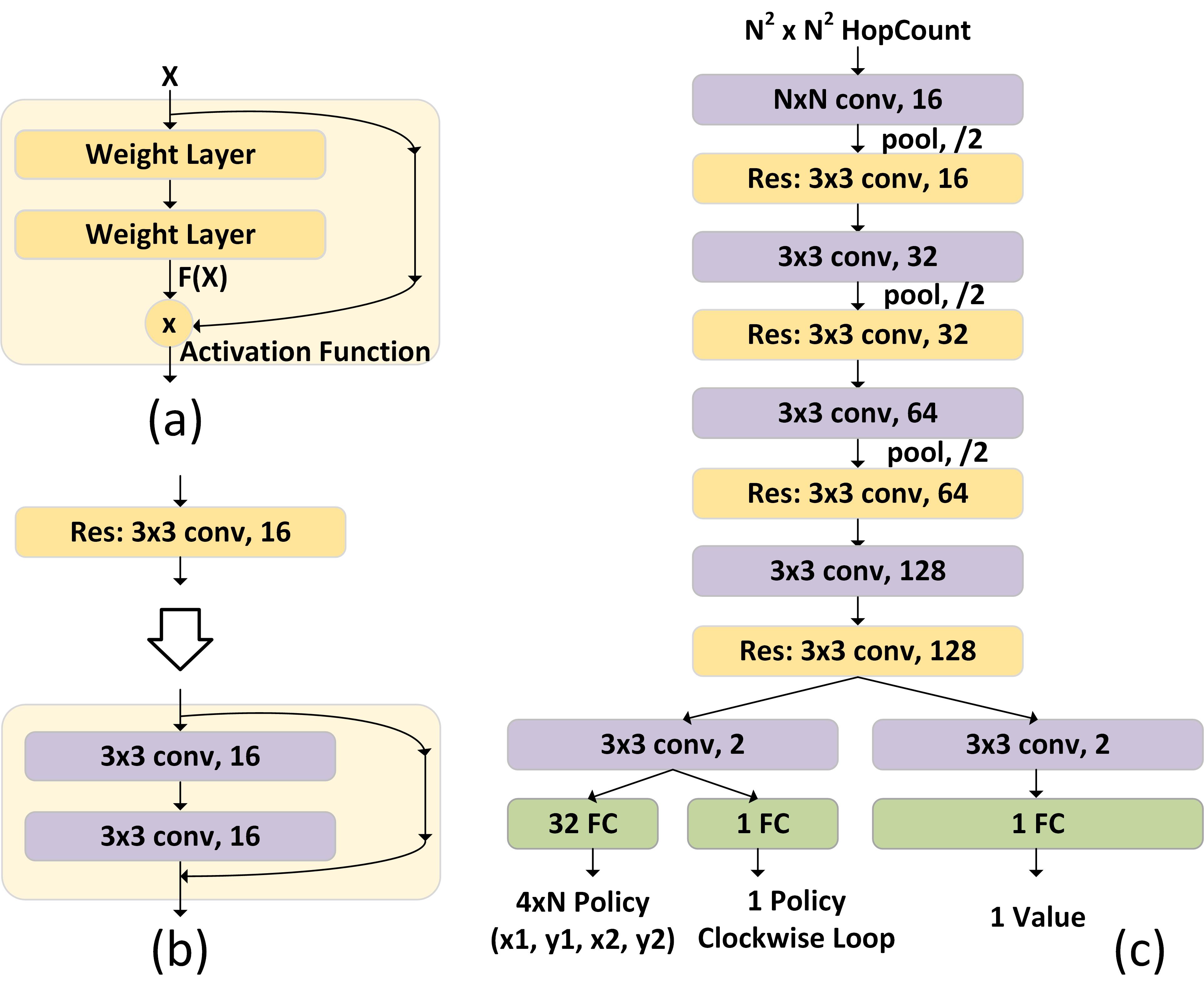}
\caption{Deep residual networks. (a) A generic building block for residual networks. (b) A building block for convolutional residual networks. (c) Proposed network.} 
\label{fig:DNN}
\end{figure}

  \textbf{Residual Neural Networks:} Sufficient network depth is essential and, in fact, leading results have used at least ten DNN layers \cite{Silver2016,Silver2017,He2016}. High network depth, however, causes overfitting for many standard DNN topologies. Residual networks offer a solution by introducing additional shortcut connections between layers that allow robust learning even with network depths of 100 or more layers. A building block for residual networks is shown in Figure \ref{fig:DNN}(a). Here, the input is $X$ and the output, after two weight layers, is $F(X)$. Notice that both $F(X)$ and $X$ (via the shortcut connection) are used as input to the activation function. This shortcut connection provides a reference for learning optimal weights and mitigates the vanishing gradient problem during back propagation \cite{He2016}. Figure \ref{fig:DNN}(b) depicts a residual box (Res) consisting of two convolutional (conv) layers. Here, the numbers 3x3 and $16$ indicate a 3x3x16 convolution kernel.  
  
  \textbf{DNN architecture:} The proposed DNN uses the two-headed architecture shown in Figure \ref{fig:DNN}(c), which learns both the policy function and the value function. This structure has been proven to reduce the amount of data required to learn the optimal policy function \cite{Silver2017}. We use convolutional layers because loop placement analysis is similar to spatial analysis in image segmentation, which performs well on convolutional neural networks. Batch normalization is used after convolutional layers to normalize the value distribution and max pooling (denoted "pool") is used after specific layers to select the most significant features. Finally, both policy and value estimates are produced at the output as the two separate heads. The policy, discussed in section \ref{representation}, has two parts: the four dimensions, $x1, y1, x2, y2$, are generated by a softmax function following a ReLU while $dir$ is generated separately using a tanh function. Tanh is used for direction as its output is between -1 and 1 whereas ReLU's output is between 0 and $\infty$. The value head uses a single convolutional layer followed by a fully connected layer, without an activation function, to predict cumulative returns.
  
\textbf{Gradients for DNN Training:} In this subsection we derive parameter gradients for the proposed DNN architecture.\footnote{Although not essential for understanding the work, this subsection provides theoretical support and increases reproducibility.} We define $\tau$ as the search process for a routerless NoC in which an agent receives a sequence of returns $\{r_t\}$ after taking actions $\{a_t\}$ from each state $\{s_t\}$. This process $\tau$ can be described a sequence of states, actions, and returns:
\begin{align}
\label{eq:tai}
  & \tau = (s_0, a_0, r_0, s_1, a_1, r_1, s_1, ...).
\end{align}

A given sequence of loops is added to the routerless NoC with probability (i.e., $\tau\sim p(\tau |\theta)$). We can then write the expected cumulative returns for one sequence as 
\begin{align}
  & \mathbb{E}_{\tau\sim p(\tau |\theta)}[r(\tau)] = \int_\tau r(\tau)p(\tau ; \theta)d\tau \\
\label{eq:Etai}
  & p(\tau ; \theta) = p(s_0)\prod_{t\geq0}\pi(a_t|s_t, \theta)P(s_{t+1, r_t}|s_t, a_t),
\end{align}
where $r(\tau)$ is a return and $\theta$ is DNN weights/parameters we want to optimize. We then differentiate the expected cumulative returns for parameter gradients
\begin{align}
\label{eq:grad}
  & \nabla \mathbb{E}_{\tau\sim p(\tau |\theta)}[r(\tau)] = \nabla_\theta\int_\tau r(\tau)p(\tau ; \theta)d\tau \\
  &\quad = \int_\tau (r(\tau)\nabla_\theta\log p(\tau ; \theta))p(\tau ; \theta)d\tau\\
\label{eq:grad1}
  &\quad = \mathbb{E}_{\tau_\theta\sim p(\tau |\theta)}[r(\tau)\nabla_\theta\log p(\tau ; \theta)].
\end{align}
Notice that transition probability $P(s_{t+1, r_t}|s_t, a_t)$ is independent of $\theta$ so we can rewrite Equation \ref{eq:grad1} as
\begin{align}
\label{eq:gradsecond}
  & \quad \mathbb{E}_{\tau_\theta\sim p(\tau |\theta)}[r(\tau)\nabla_\theta\log p(\tau ; \theta)]\\
  &\quad = \mathbb{E}_{\tau_\theta\sim p(\tau |\theta)}[r(\tau)\nabla_\theta\Sigma\log \pi(a_t ; s_t, \theta)]\\
\label{eq:gradsecond0}
  &\quad \approx \sum_{t\geq0}r(\tau)\nabla_\theta\log \pi(a_t ; s_t, \theta).
\end{align}
The equation \ref{eq:gradsecond0} gradient is proportional to raw returns. We rewrite equation \ref{eq:gradsecond0} to minimize the difference between predictions and real values as
\begin{align}
\label{eq:gradsecond2}
  & \nabla_\theta \mathbb{E}_{\tau\sim p(\tau |\theta)}[r(\tau)] \approx \sum_{t\geq0}A_t\nabla_\theta\log \pi(a_t ; s_t, \theta)\\
  & A_t = \sum_{t'>t}\gamma^{\ t'-t}r_{t'}-V(s_t;\theta),
\end{align}
where $A_t$ is the advantage/difference between the predictions and real values. This approach is also known as advantage actor-critic learning where the actor and the critic represent the policy function and value function, respectively \cite{RLbook}. In a two-headed DNN, $\theta$ consists of $\theta_\pi$ and $\theta_v$ for the policy function and the value function, respectively. The gradients for the two parameter sets are then given as
\begin{align}
  & d\theta_\pi = (\sum_{t'>t}\gamma^{\ t'-t}r_{t'}-V(s_t;\theta_v)) \nabla_{\theta_\pi}\log \pi(a_t ; s_t, \theta_\pi)\\
  & d\theta_v = \nabla_{\theta_v}(\sum_{t'>t}\gamma^{\ t'-t}r_{t'}-V(s_t;\theta_v))^2.
\end{align}
The whole training procedure repeats the following equations
\begin{align}
\label{eq:gradsecond3}
  & \theta_\pi = \theta_\pi + \gamma*d\theta_\pi\\
\label{eq:gradsecond4}
  & \theta_v = \theta_v + c*\gamma*\theta_v,
\end{align}
where $\gamma$ is a learning rate and $c$ is a constant.

\subsection{Routerless NoC Design Exploration}
  An efficient approach for design space exploration is essential for routerless NoC design due to the immense design space. Deep reinforcement learning approaches are therefore well-suited for this challenge as they can leverage recorded states while learning. Some work uses experience replay, which guides actions using random samples. These random samples are useful throughout the entire learning process, so improve collected state efficiency \cite{Mnih2013}, but break the correlation between states. Another approach is the Monte Carlo tree search (MCTS), which is more closely correlated to human learning behavior based on experience. MCTS stores previously seen routerless NoC configurations as nodes in a tree structure. Each node is then labeled with the expected returns for exploration starting from that node. As a result, MCTS can provide additional insight during state exploration and help narrow the scope of exploration to a few promising branches \cite{Silver2016} to efficiently learn optimal loop placement.
  
  In our implementation, each node $s$ in the tree represents a previously seen routerless NoC and each edge represents an additional loop. Additionally, each node $s$ stores a set of statistics: $\overline{V}(s_{next})$, $P(a_i; s)$, and $N(a_i; s)$. $\overline{V}(s_{next})$ is the mean cumulative return from $s_{next}$ and is used to approximate the value function $V^\pi(s_{next})$. $P(a_i; s)$ is the prior probability of taking action $a_i$ based on $\pi(a=a_i; s)$. Lastly, $N(a_i; s)$ is the visit count, representing the number of times $a_i$ was selected at $s$. Exploration starts from state $s$, then selects the best action $a^*$ based on expected exploration returns given by 
\begin{align}
\label{eq:a_best}
  & a^* = arg\max_{a_i}(U(s, a_i) + \overline{V}(s_{next}))\\
\label{eq:a_best2}
  & U(s, a_i) = c*P(a_i; s)\frac{\sqrt{\sum_jN(a_j; s)}}{1+N(a_i; s)},
\end{align} 
where $U(s,a_i)$ is the upper confidence bound and $c$ is a constant \cite{Rosin2011}. The first term in Equation \ref{eq:a_best} encourages broad exploration while the second emphasizes fine-grained exploitation. At the start, $N(a_i; s)$ and $\overline{V}(s_{next})$ are similar for most routerless NoCs so exploration is guided by $P(a_i; s)=\pi({a=a_i;s})$. Reliance upon DNN policy decreases with time due to an increasing $N(a_i; s)$, which causes the search to asymptotically prefer actions/branches with high mean returns \cite{Silver2017}. Search is augmented by an $\epsilon$-greedy factor where the best action is ignored with probability $\epsilon$ to further balance exploration and exploitation.

\begin{figure}[ht]
\centering
\includegraphics[width=0.48\textwidth]{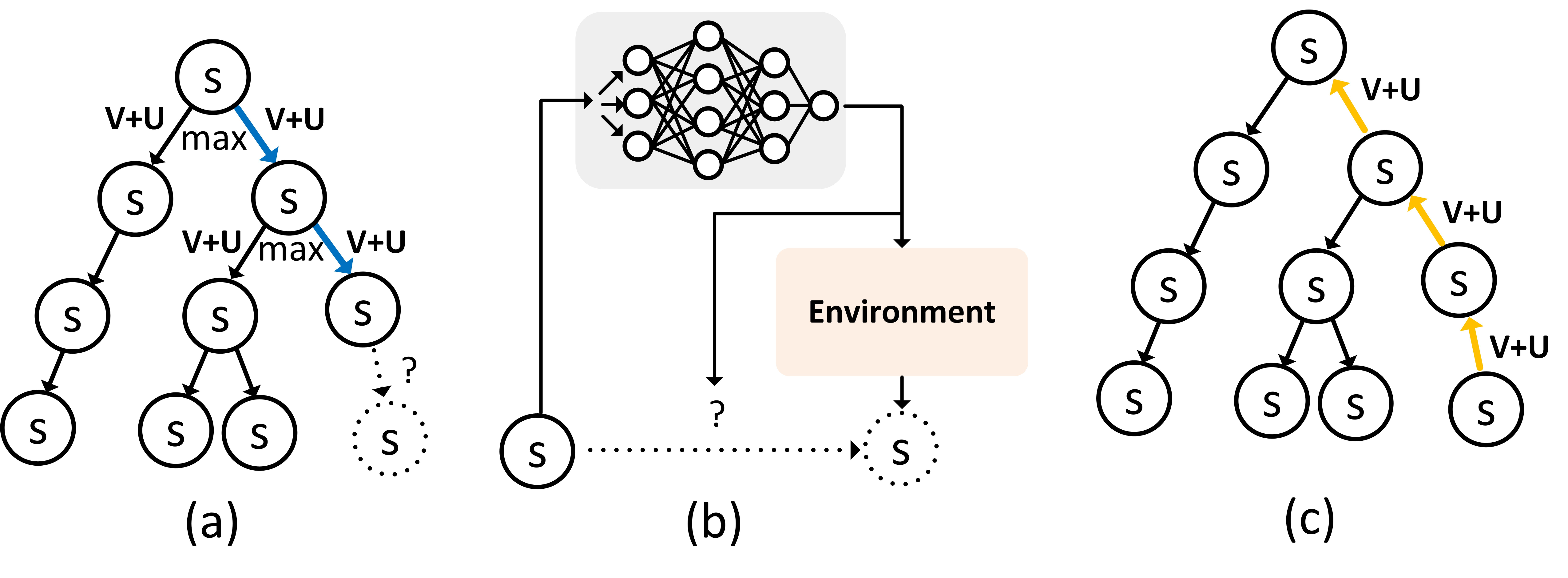}
\caption{Monte Carlo tree search. (a) Search. (b) Expansion+evaluation using DNN. (c) Backup. }
\label{fig:MCTS}
\end{figure}

  There are three phases to the MCTS algorithm shown in Figure \ref{fig:MCTS}: search, expansion+evaluation, and backup. (1) Search: an agent selects the optimal action (loop placement) by either following Equation \ref{eq:a_best} with probability $1-\epsilon$ or using a greedy search with probability $\epsilon$. Algorithm \ref{alg:greedy} details the greedy search that evaluates the benefit from adding various loops and selects the loop with the highest benefit. $CheckCount()$ returns the total number of nodes that can communicate after adding a loop with diagonal nodes at $(x1, y1)$ and $(x2, y2)$. Next, the $Imprv()$ function returns the preferred loop direction based on the average hop count improvement. The tree is traversed until reaching a leaf node (NoC configuration) without any children (further developed NoCs). (2) Expansion+evaluation: the leaf state is evaluated using the DNN to determine an action for rollout/expansion. Here, $\pi(a=a_i; s)$ is copied, then later used to update $P(a_i; s)$ in Equation \ref{eq:a_best2}. A new edge is then created between $s$ and $s_{next}$ where $s_{next}$ represents the routerless NoC after adding the loop to $s$. (3) Backup: After the final cumulative returns are calculated, statistics for the traversed edges are propagated backwards through the tree. Specifically, $\overline{V}(s_{next})$, $P(a_i; s)$, and $N(s, a_i)$ are all updated.
   
\begin{algorithm}
\caption{Greedy Search}\label{alg:greedy}
\begin{algorithmic}[1]
\State Initialization: bestLoop = [0, 0, 0, 0], bestCount = 0, bestImprv = 0, and dir = 0
\For{x1 = 1;+1;N}
	\For{y1 = 1:+1;N}
		\For{x2 = x1+1:+1;N}
        	\For{y2 = y1+1:+1;N}
            	\State count = $CheckCount(x1, y1, x2, y2)$
                \If{count > bestCount}
                	\State bestCount = count
                    \State bestLoop = [x1, y1, x2, y2]
					\State bestImpv, dir = $Imprv(x1, y1, x2, y2)$
                \ElsIf{return == bestCount}
                	\State imprv', dre' = $Imprv(x1, y1, x2, y2)$
                    \If {imprv' > bestImprv}
                    	\State bestLoop = [x1, y1, x2, y2]
                        \State bestImprv = imprv'
                        \State dir = dir'                
                    \EndIf
                \EndIf                
            \EndFor
       \EndFor
	\EndFor
\EndFor
\State \Return bestRing, dir
\end{algorithmic}
\end{algorithm}

\subsection{Multi-threaded Learning}
\begin{figure}[ht]
\centering
\includegraphics[width=0.45\textwidth]{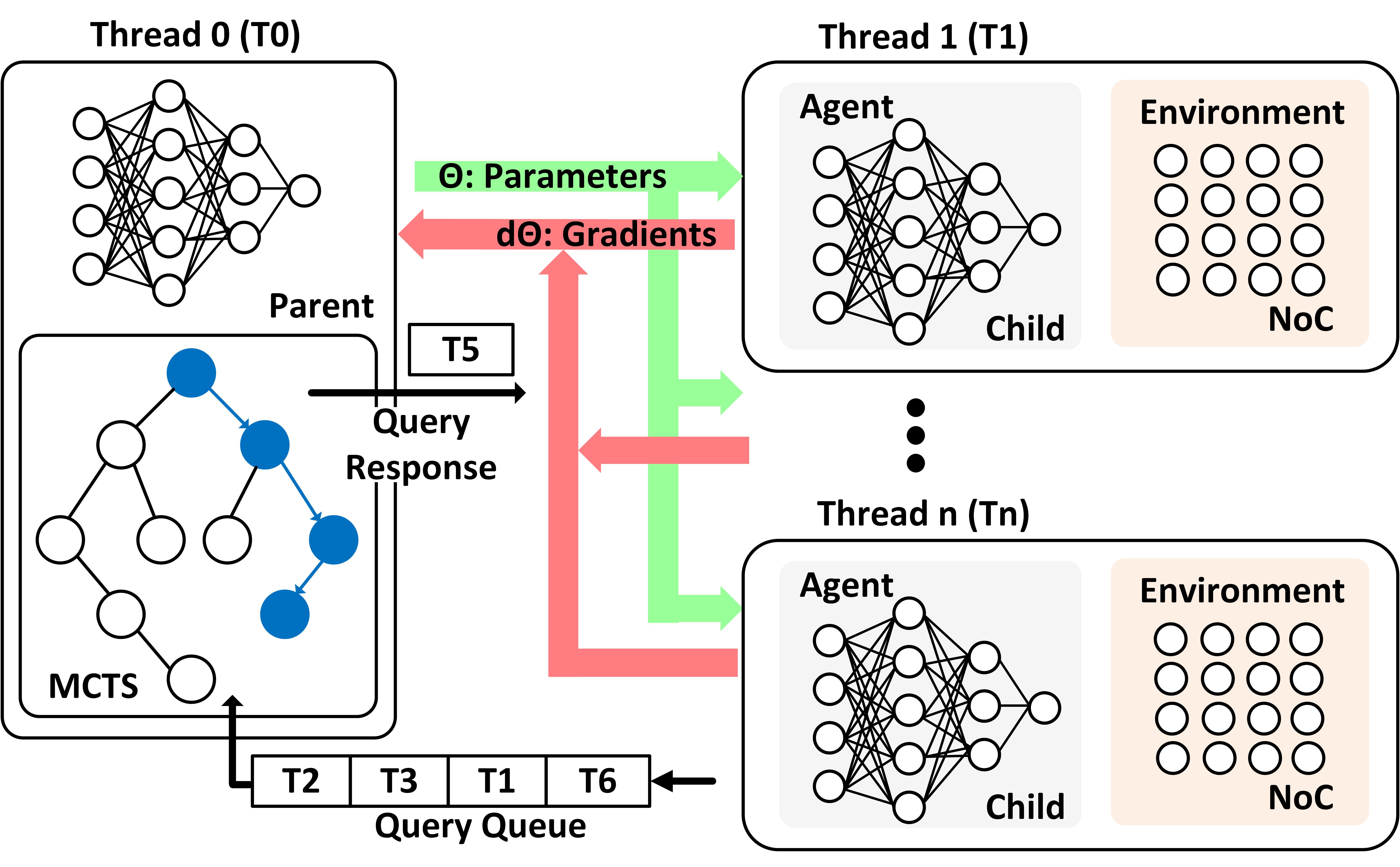}
\caption{Multi-threaded framework.}
\label{fig:framework}
\end{figure}

 The framework incorporates a multi-threaded approach, in which many threads independently explore the design space while collaboratively updating global parameters. This facilitates efficient exploration for optimal routerless NoC configurations \cite{Mnih2016}. Figure \ref{fig:framework} depicts the proposed framework with multi-threaded exploration. At the start, thread 0 creates a parent DNN with initial weights/parameters $\theta$, then creates many child threads (1 to n) that create their own child DNNs, each of which acts as an individual learning agent. The parent thread sends DNN parameters to child threads and receives parameter gradients from child threads. This multi-threaded approach stabilizes convergence by averaging both large gradients and small gradients during training \cite{Mnih2016}. The parent thread additionally maintains a search tree that records past child thread actions for each MCTS query.

\section{Methodology}

  We evaluate the proposed deep reinforcement learning (DRL) routerless design against the previous state-of-the-art routerless design (REC) \cite{RL} and several mesh configurations. All simulations use Gem5 with Garnet2.0 for cycle-accurate simulation \cite{gem5}. For synthetic workloads, we test uniform random, tornado, bit complement, bit rotation, shuffle, and transpose traffic patterns. Performance statistics are collected for 100,000 cycles across a range of injection rates, starting from 0.005 flits/node/cycle and incremented by 0.005 flits/node/cycle until the network saturates. Results for PARSEC are collected after benchmarks are run to completion with either sim-large or sim-medium input sizes.\footnote{Several workloads exhibit compatibility issues with our branch of Gem5, but we include all workloads that execute successfully.} Power and area estimations are based on Verilog post-synthesis simulation, following a similar VLSI design flow in REC that synthesizes the Verilog implementation in Synopsys Design Compiler and conducts Place \& Route in Cadence Encounter under 15nm NanGate FreePDK15 Open Cell Library \cite{Encounter}.

  We regard node overlapping as a more appropriate measure than link overlapping (i.e., the number of links between adjacent nodes) for manufacturing constraints. For fair comparison, we use the node overlapping generated by the algorithm in REC as a reference. Node overlapping values are given in Table \ref{tab:NOCtab}. Loop configurations for DRL are generated using our proposed framework, described in Section 4, with the desired node overlapping.

  For synthetic and PARSEC workloads, REC and DRL variants use identical configurations for all other parameters, matching prior testing \cite{RL} for comparable results. Results nevertheless differ slightly due to differences between Gem5 and Synfull \cite{Synfull}, used in REC testing. In REC and DRL, each input link is attached to a flit-sized buffer with 128-bit link width. Packet injection and forwarding can each finish in a single cycle up to 4.3 GHz. For all mesh simulations, we use a standard two-cycle router delay in our baseline (Mesh-2). We additionally test an optimized one-cycle delay router (Mesh-1) and, in PARSEC workloads, an "ideal" router with zero router delay (Mesh-0) leaving only link and contention delays. These mesh configurations all use 256-bit links, 2 VCs per link, and 4-flit input buffers. Packets are categorized into control and data packets, with 8 bytes and 72 bytes, respectively. The number of flits per packet is then given as packet size divided by link width. Therefore, in REC and DRL simulations, control packets are 1 flit and data packets are 5 flits. Similarly, in mesh simulations, control packets are 1 flit while data packets are 3 flits. For PARSEC workloads, L1D and L1I cache sizes are set to 32 KB with 4-way associativity and L2 cache is set to 128 KB with 8-way associativity. Link delay is set to one cycle per hop for all tests.

\begin{table}
    \centering
    \caption{NoC Node Overlap and Loop Count}
    \small
    \label{tab:NOCtab}
        \begin{tabular}{|c|c|c|c|c|}
            \hline
            NoC &
            \multicolumn{2}{|c|}{Node Overlapping } &
            \multicolumn{2}{|c|}{Loop Count } \\
            \cline{2-5}
            Size & 
            \multicolumn{2}{|c|}{REC \& DRL } &
            REC & DRL \\
            \hline
            4x4 & 
            \multicolumn{2}{|c|}{ 6 } & 
            10 & 10 \\
            \hline
            6x6 & 
            \multicolumn{2}{|c|}{ 10 } &
            24 & 27 \\
            \hline
            8x8 &
            \multicolumn{2}{|c|}{ 14 } &
            44 & 52 \\
            \hline
            10x10 & 
            \multicolumn{2}{|c|}{ 18 } & 
            70 & 74 \\
            \hline
		\end{tabular}
\end{table}

\section{Results \& Analysis}
\subsection{Design Space Exploration}
\begin{figure}[ht]
\centering
\includegraphics[width=0.45\textwidth]{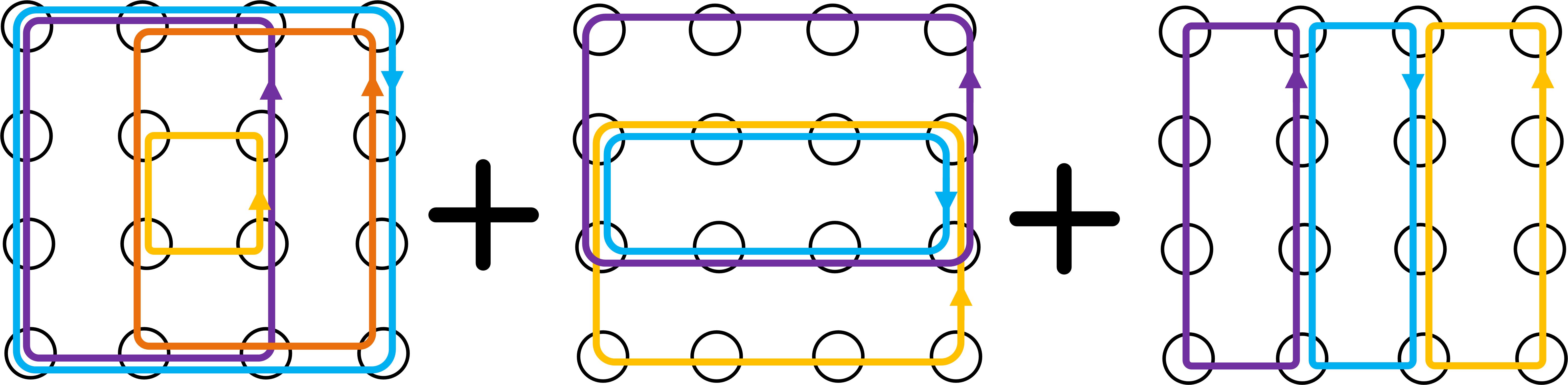}
\caption{A 4x4 NoC topology generated by DRL.}
\label{fig:DRL}
\end{figure}

The agent starts without $a$ $priori$ experience or training data. Over time, as the search tree is constructed, the agent begins to explore more useful loop configurations, which provide increased performance. Configurations satisfying design criteria can be found in seconds and minutes for 4x4 and 10x10 NoCs, respectively. Figure \ref{fig:DRL} illustrates a 4x4 DRL design. The generated topology is interestingly structured similarly to REC \cite{RL}, using only rectangular loops, but replaces one inner loop with a larger loop and explores different loop directions. The resulting topology is by no means arbitrary and, in a 4x4 NoC, is completely symmetric and far more regular than IMR. We observe similar structure for 8x8 and 10x10 topologies, but omit these due to space constraints.

\subsection{Synthetic Workloads}
\begin{figure*}
 \center
  \includegraphics[width=\textwidth]{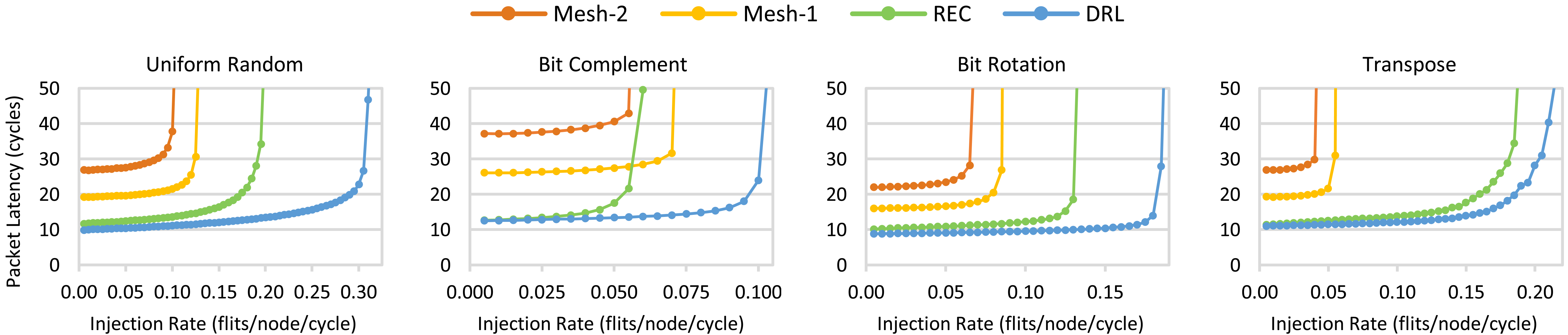}
  \caption{Average packet latency for synthetic workloads in 10x10 NoC.}
  \label{fig:Synth}
\end{figure*}

\textbf{Packet Latency:} Figure \ref{fig:Synth} plots the average packet latency of four synthetic workloads for a 10x10 NoC. Tornado and shuffle are not shown as their trends are very similar to bit rotation. Zero-load packet latency for DRL is the lowest in all workloads. For example, with uniform random traffic, zero-load packet latency is 9.89, 11.67, 19.24, and 26.85 cycles for DRL, REC, Mesh-1, and Mesh-2, respectively, corresponding to a 15.2\%, 48.6\%, and 63.2\% latency reduction by DRL. Across all workloads, DRL reduces zero-load packet latency by 1.07x, 1.48x and 1.62x compared with REC, Mesh-1, and Mesh-2, respectively. This improvement for both REC and DRL over Mesh configurations results from reduced per hop latency (one cycle). DRL improves over REC due to additional connectivity and better loop placement. Observing Table \ref{tab:NOCtab}, in a 10x10 NoC, DRL provides four additional paths that tend to connect more nodes due to a more effective search process. 

\textbf{Throughput:} DRL provides substantial throughput improvements for all traffic patterns. For uniform traffic, throughput is approximately 0.1, 0.125, 0.195, and 0.305 for Mesh-2, Mesh-1, REC, and DRL, respectively. Notably, in transpose, DRL improves throughput by 208.3\% and 146.7\% compared with Mesh-2 and Mesh-1. Even in bit complement where mesh configurations perform similarly to REC, DRL still provides a 42.8\% improvement over Mesh-1. Overall, DRL improves throughput by 3.25x, 2.51x, and 1.47x compared with Mesh-2, Mesh-1, and REC, respectively. Again, additional loops with greater connectivity in DRL allow a greater throughput compared with REC. Furthermore, improved path diversity provided by these additional loops allows much higher throughput compared with mesh configurations.

\subsection{PARSEC Workloads}

\begin{figure}
 \center
  \includegraphics[width=0.48\textwidth]{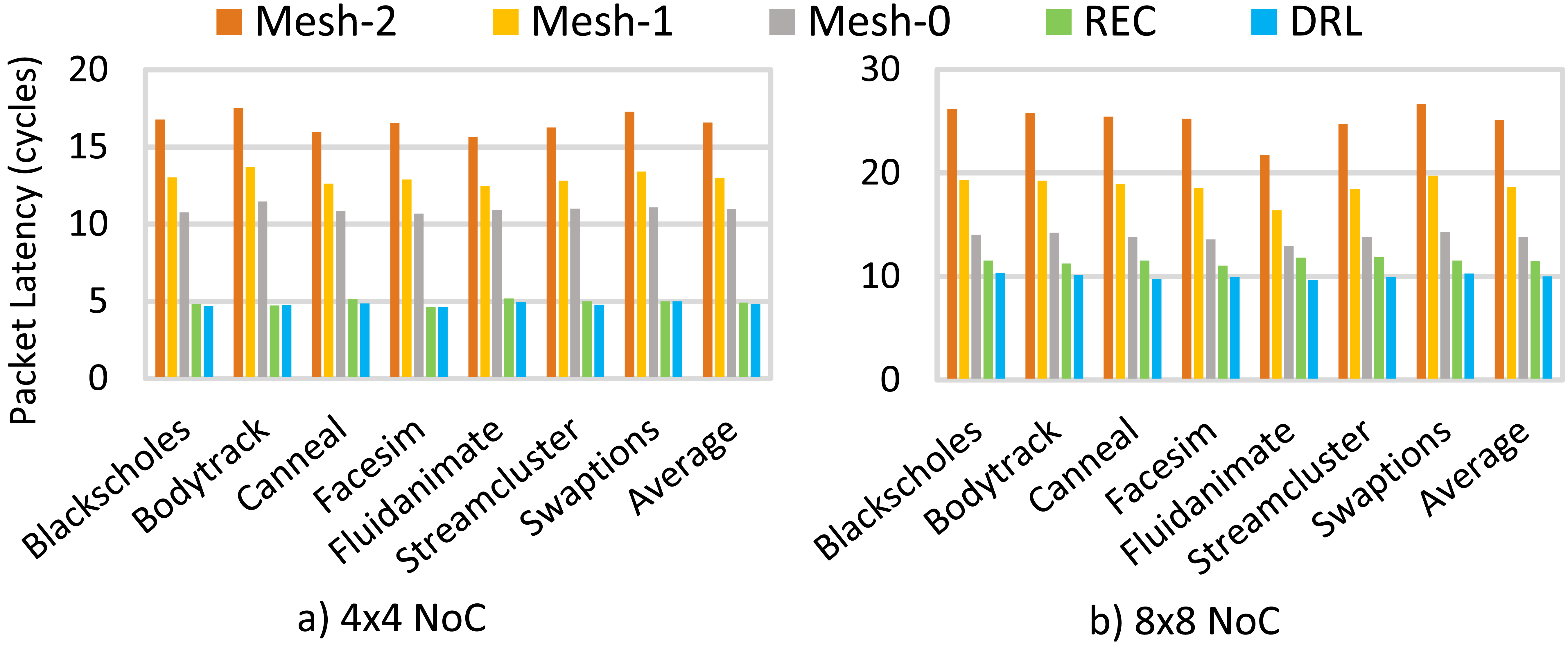}
  \caption{Packet latency for PARSEC workloads.}
  \label{fig:PARSEC}
\end{figure}

We compare real-world application performance for REC, DRL, and three mesh configurations with a set of PARSEC benchmarks. We generate Mesh-0 results by artificially reducing packet latency by the hop count for every recorded flit since such a configuration is difficult to simulate otherwise. As a result, performance is close to but slightly worse than a truly "ideal" zero-cycle-router mesh. NoC sizes of 4x4 and 8x8 are evaluated using Gem5.
 
\textbf{Packet Latency:} As shown in Figure \ref{fig:PARSEC}, for the 4x4 network, variations in loop configuration are relatively small, being heavily influenced by full-connectivity requirements. Nevertheless, in the 4x4 NoC, DRL improves performance over REC in all but two applications where performance is similar. For example, DRL reduces packet latency by 4.7\% in fluidanimate compared with REC. Improvements over mesh configurations for fluidanimate are greater with a 68.5\%, 60.4\%, and 54.9\% improvement compared with Mesh-2, Mesh-1, and Mesh-0. On average, DRL reduces packet latency by 70.7\%, 62.8\%, 56.1\%, and 2.6\% compared with Mesh-2, Mesh-1, Mesh-0, and REC, respectively.

DRL improvements are substantial in the 8x8 NoC as DRL can explore a larger loop configuration design space. For example, in fluidanimate, average packet latency is 21.7, 16.4, 12.9, 11.8, and 9.7 in  Mesh-2, Mesh-1, Mesh-0, REC, and DRL, respectively. This corresponds to a 55.6\%, 41.0\%, 25.3\%, and 18.2\% improvement for DRL compared with Mesh-2, Mesh-1, Mesh-0, and REC. On average, DRL reduces packet latency by 60.0\%, 46.2\%, 27.7\%, and 13.5\% compared with Mesh-2, Mesh-1, Mesh-0, and REC, respectively. 

\begin{figure}[ht]
	\centering
	\includegraphics[width=0.48\textwidth]{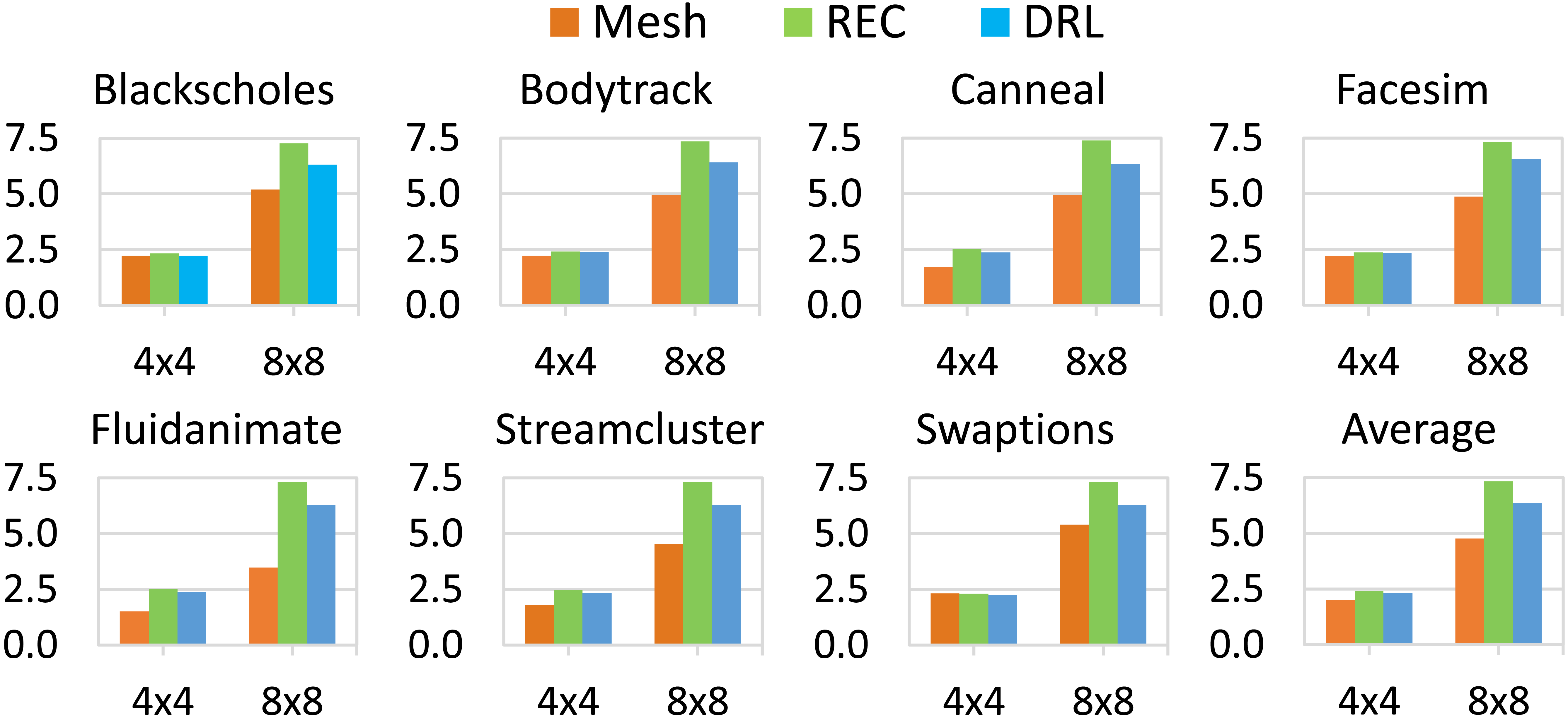}
	\caption{Average hop count for PARSEC workloads.}
\label{fig:PARSEC_hops}
\end{figure}

\textbf{Hop Count:} Figure \ref{fig:PARSEC_hops} compares the average hop count for REC, DRL, and Mesh-2 for 4x4 and 8x8 NoCs. Only Mesh-2 is considered as differences in hop count are negligible between mesh configurations (they mainly differ in per-hop delay). For 4x4 networks, REC and DRL loop configurations are relatively similar so improvements are limited, but DRL still provides some improvement in all workloads compared with REC. In streamcluster, average hop count is 1.79, 2.48, and 2.34 for mesh, REC, and DRL, respectively. On average, DRL hop count is 22.4\% higher than mesh and 3.8\% less than REC. For larger network sizes, we again observe the benefit from increased flexibility in loop configuration that DRL exploits. This optimization allows more loops to be generated, decreasing average hop count compared with REC by a minimum of 12.7\% for bodytrack and a maximum of 14.3\% in fluidanimate. On average, hop count for DRL is 13.7\% less than REC and 35.7\% higher than mesh.

\textbf{Execution Time:} Execution times for 8x8 PARSEC workloads are given in Table \ref{tab:ExecTable}. Reductions in hop count and packet latency may not necessarily translate to reduced execution time as applications may be insensitive to NoC performance (notably streamcluster). Nevertheless, in fluidanimate, a NoC sensitive workload, DRL reduces execution time by 30.7\% over Mesh-2, 16.4\% over Mesh-1, and 3.17\% over REC. Overall, DRL provides the smallest execution time for every workload and, on average, DRL's execution is 13.3\% faster than Mesh-2, 7.1\% faster than Mesh-1, and 1.0\% faster than REC. Note that NoC traffic for PARSEC workloads is known to be light, so the significant throughput advantage of DRL over mesh and REC (as seen in Figure \ref{fig:Synth}) is not fully demonstrated here.

\begin{table}
    \centering
    \caption{8x8 PARSEC workload execution time (ms).}
    \small
    \label{tab:ExecTable}
        \begin{tabular}{|c|c|c|c|c|}
            \hline
            Workload &
            \multicolumn{4}{|c|}{ NoC Type } \\
            \cline{2-5}
             & Mesh-2 & Mesh-1 & REC & DRL \\
            \hline
            Blackscholes & 4.4 & 4.2 & 4.0 & 4.0 \\
            \hline
            Bodytrack & 5.4 & 5.3 & 5.1 & 5.1 \\
            \hline
            Canneal & 7.1 & 6.4 & 6.1 & 6.0 \\
            \hline
            Facesim & 626.0 & 587.0 & 515.2 & 512.3 \\
            \hline
            Fluidanimate & 35.3 & 29.2 & 25.2 & 24.4 \\
            \hline
            Streamcluster & 11.0 & 11.0 & 11.0 & 11.0 \\
            \hline
		\end{tabular}
\end{table}

\subsection{Power}

\begin{figure}[ht]
	\centering
	\includegraphics[width=0.48\textwidth]{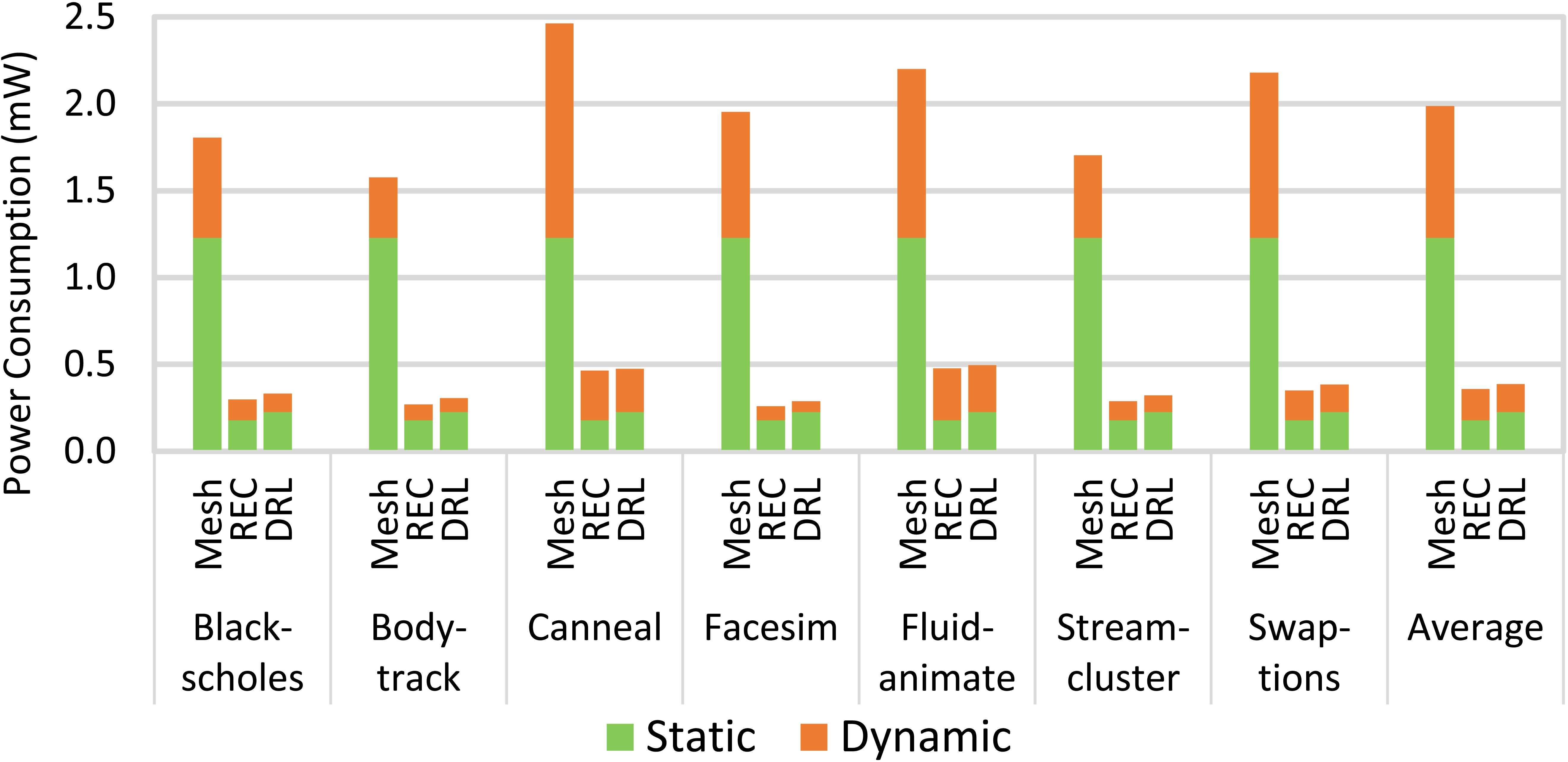}
	\caption{Power consumption for PARSEC workloads.}
	\label{fig:Power}
\end{figure}

  Figure \ref{fig:Power} compares the power consumption of REC, DRL, and Mesh (Mesh-2) across PARSEC workloads. Results are generated after Place \& Route in Cadence Encounter under 15nm technology node \cite{Encounter}. The global activity factor is estimated with the link utilization statistics from Gem5 simulations. Verilog implementation uses this activity factor for several sets of parameters to appropriately represent average power across the 8x8 NoC. Values are reported using weighted averages. A clock frequency of 2.0 GHz is used, comparable to commercial many-core processors.
  
  Static power is 0.18mW for REC and 0.23mW for DRL, both of which are considerably lower than the 1.23 mW of Mesh. The slight increased from REC to DRL is expected as resources required to support the additional links, such as loop selection and buffers, will scale relatively linearly. Other resources, including ejection buffers, are constant in all configurations as a similar number will satisfy the equivalent traffic \cite{RL}. Dynamic power is the lowest for DRL due to improved resource utilization, leading to lower global activity factors and fewer active cycles compared with REC. As a result, DRL has lower dynamic power than REC in all workloads. DRL also provides significant savings over mesh due to reduced routing logic and fewer buffers. 
  On average, dynamic power for DRL is 80.8\% and 11.7\% less than Mesh and REC, respectively.

\subsection{Area}

\begin{figure}[ht]
	\centering
	\includegraphics[width=0.46\textwidth]{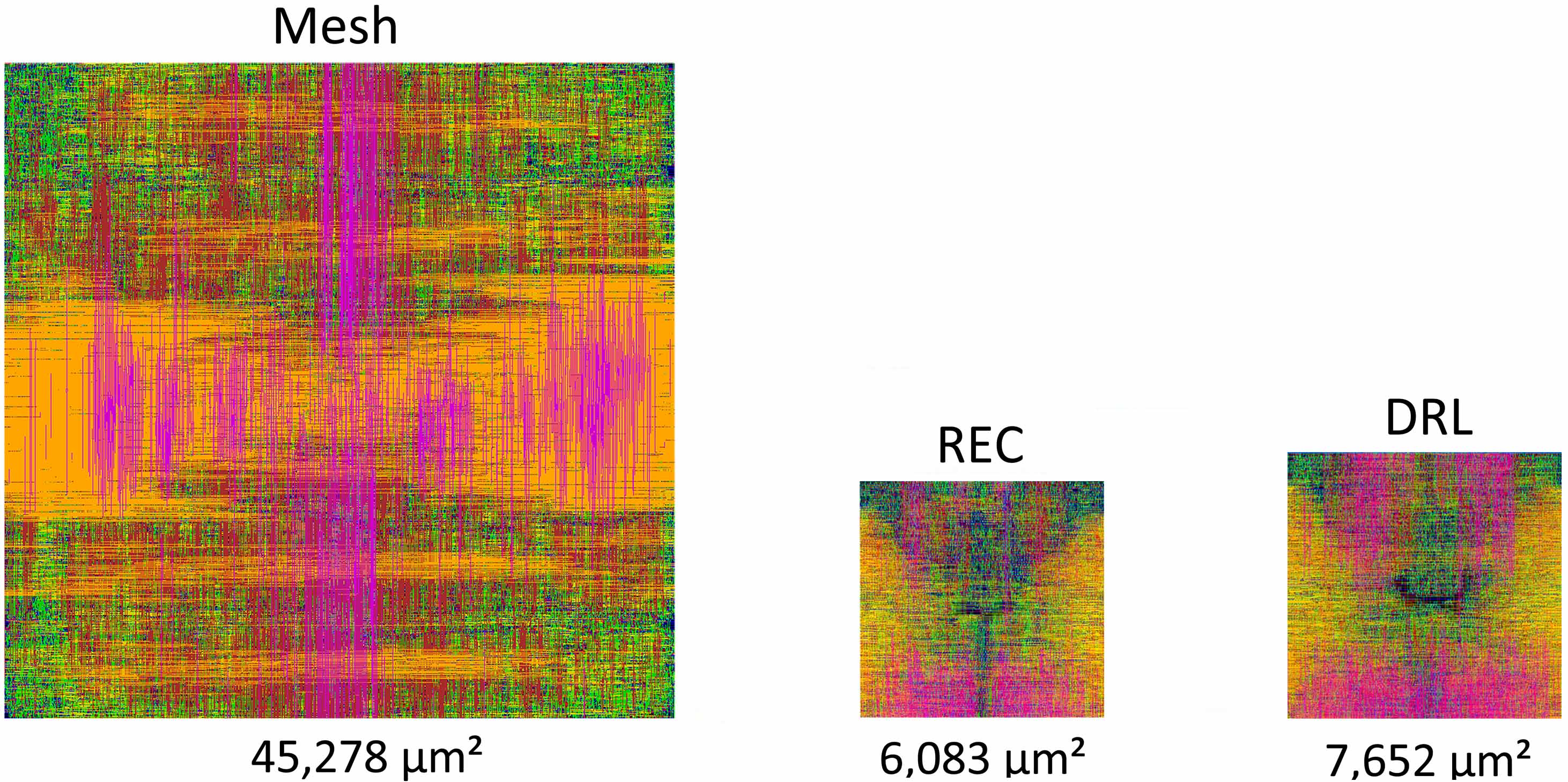}
	\caption{Area comparison (after P\&R).}
	\label{fig:Area}
\end{figure}

  Figure \ref{fig:Area} compares the interface area for REC, DRL, and Mesh (Mesh-2) configurations. Area values in the figure are given using weighted averages to represent all nodes in an 8x8 NoC. REC has the smallest area at 6,083 $\mu m^2$ as there are just 10.4 loops per node on average. The area for DRL is a bit larger at 7,652 $\mu m^2$ due to an increased average of 13.3 loops per node. Finally, the area for mesh is much higher at 45,278 $\mu m^2$. This difference is mainly attributed to the ability of routerless NoCs to avoid using crossbars and virtual channels. 
  Note that the above area results for REC and DRL have already included the small look-up table at source. The table is needed to identify which loop to use for each destination (if multiple loops are connected), but each entry has only a few bits \cite{RL}. Precisely, the area for the table and related circuitry is 443 $\mu m^2$, equivalent to only 0.9\% of the Mesh router (power is 0.028mW or 1.13\% of Mesh).

  We have also evaluated the additional repeaters necessary to support DRL. The total repeater area is 0.159 $mm^2$, so the additional overhead for DRL compared to REC represents just 1.1\% of Mesh.
  
\subsection{Discussion}

\textbf{Power and Area Overhead:} DRL primarily improves performance over REC due to more effective loop configurations. Specifically, DRL generates NoC configurations with a higher \textit{average} loop count per node (but still within the same \textit{maximum} node overlapping as REC), thus having more overall loops and links. These additional resources, however, actually allow better utilization for other interface resources, resulting in lower average dynamic power. The additional area overhead is likewise expected. 

In both power and area analysis, we assume that each node interface uses the exact number of loops generated by REC and DRL. This gives an advantage to REC by assuming that fewer loops lead to smaller interface area. In practice, to reduce design and verification efforts, all the nodes in a routerless NoC will likely use the same interface determined by the node overlapping cap, and simply leave unused loop ports idle. In that case, the static power and area for REC and DRL will be the same due to equal node overlapping.

\textbf{Comparison with IMR}: REC has previously been shown to improve over IMR in all aspects \cite{RL}. In synthetic testing, REC achieves an average 1.25x reduction in zero-load packet latency and a 1.61x improvement in throughput over IMR. Similarly, in real benchmarks, REC achieves a 41.2\% reduction in average latency. Both static and dynamic power are also significantly lower in REC due to reduced buffer requirements and more efficient wire utilization. Finally, REC area is just 6,083 $\mu m^2$ while IMR area is 20,930 $\mu m^2$, corresponding to a 2.4x increase. Comparisons between REC and DRL were therefore the primary focus in previous subsections since REC better represents the current state-of-the-art in routerless NoCs. The large gap between IMR and REC also illustrates that traditional design space search (e.g., genetic algorithm in IMR) is far from sufficient, which calls for more intelligent search strategies.

\textbf{Reliability:} Reliability concerns for routerless NoC stem from the limited path diversity since wiring constraints restrict the total number of loops. For a given node overlapping, DRL designs provide more loops and thus more paths between nodes as more nodes approach the node overlapping cap. In the 8x8 NoC, there are, on average, 2.77 paths between any two nodes in REC. This increases to 3.79 paths, on average, between any two nodes in DRL. DRL can therefore tolerate more link failures before the NoC fails.

\begin{figure}[ht]
	\centering
	\includegraphics[width=0.48\textwidth]{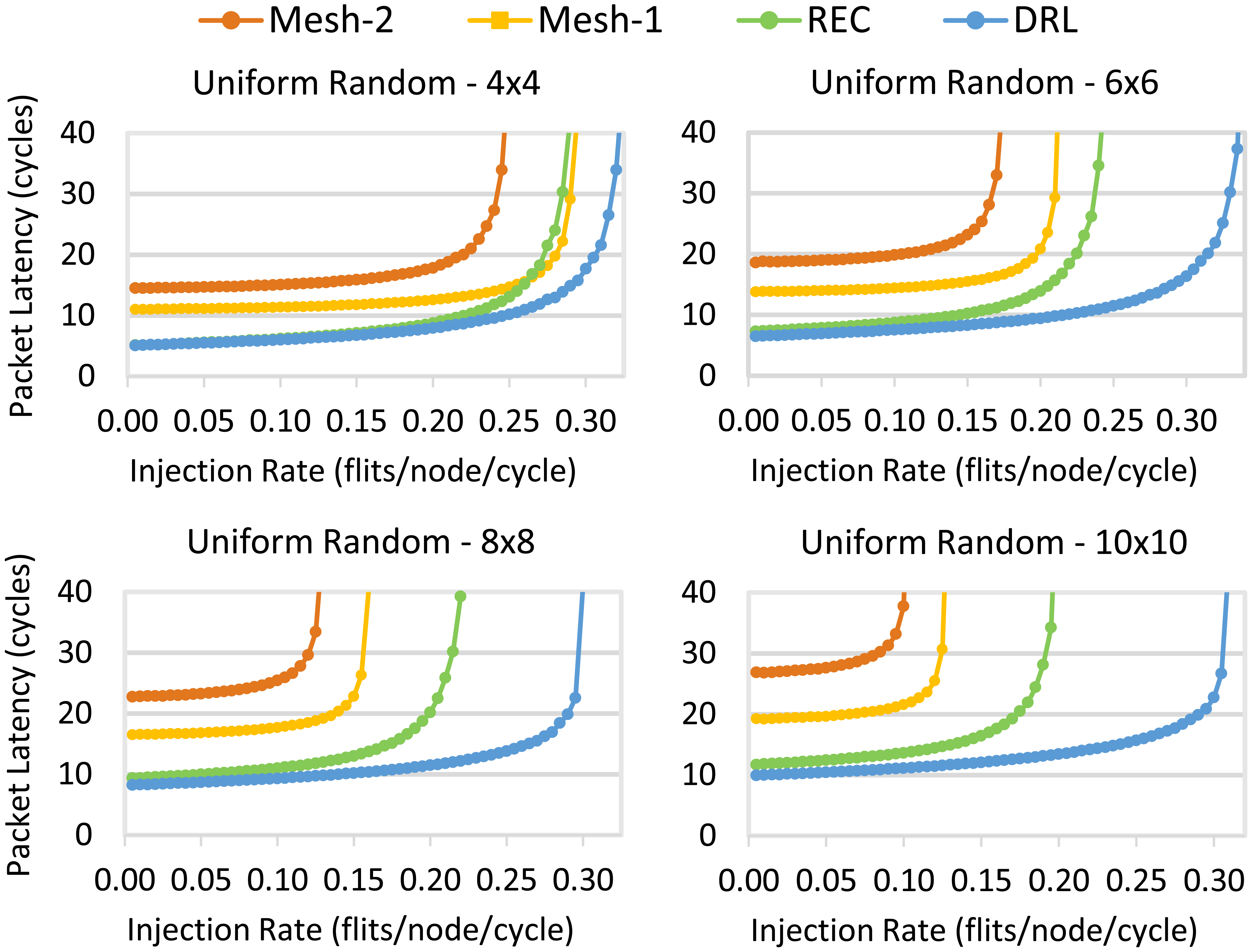}
	\caption{Synthetic Scaling for NoC Configurations.}
	\label{fig:scaling}
\end{figure}

\textbf{Scalability:} DRL scales very well compared with both REC and mesh configurations. For PARSEC workloads, shown in Figure \ref{fig:PARSEC}, the difference in packet latency between REC and DRL increases from an 2.6\% improvement in the 4x4 NoC to a 13.5\% improvement in the 8x8 NoC. Average hop count, shown in Figure \ref{fig:PARSEC_hops}, exhibits a similar trend. DRL improves average hop count by 3.8\% in a 4x4 NoC and 13.7\% in an 8x8 NoC. Scaling improvements are more evident in synthetic workloads. Figure \ref{fig:scaling}, for example, shows scaling results for 4x4 to 10x10 NoC sizes with uniform random workloads. Note that the same axis values are used for all NoC sizes to emphasize scaling performance. Whereas REC throughput decreases from 0.285 flits/node/cycle to 0.195 flits/node/cycle, corresponding to a 31.6\% decrease, the throughput for DRL only changes slightly from 0.32 to 0.305 flits/node/cycle, corresponding to a 4.7\% reduction. Increasing the NoC size allows more flexibility in loop exploration, and thus more effective use of wiring resources for a given node overlapping constraint. Additionally, loop designs for $N \times M$ NoCs using DRL is straightforward to implement, only requiring modifications to the DNN for dimension sizes.

\subsection{Broad Applicability} \label{BroadApp}
Routerless NoC design represents just one possible application for the framework presented in this paper. This framework, with modifications to state/action representations, could also be applied to router-based NoC designs. Specifically, one related application is in 3-D NoCs where higher dimensionality encourages novel design techniques. Prior work has explored small-world router-based designs \cite{3DNoC2015, Energy3DNoC} using a relatively limited learning-based approach. The design space exploration would be more effective if our framework is used. Specifically, state representation using hop count remains compatible with the current DNN structure by concatenating matrices for each 2D layer. Actions can involve adding links between nodes in the same layer (intra-layer links) or different layers (inter-layer links). One DNN can be used for each action type to achieve an efficient deep reinforcement learning process with a smaller design space. A significant advantage of our framework is that strict constraints can be enforced on link addition, such as 3-D distance, to meet timing/manufacturing capabilities.

The proposed framework can also be generalized to apply to other research problems related to NoCs. While detailed exploration is beyond the scope of this paper, we briefly mention here a few promising examples that can benefit from our framework. 
One example is to exploit the underutilized wiring resources in silicon interposer \cite{Interposer2014, Interposer2016} and use the framework to explore better ways of connecting CPU cores and stacked memories. The framework could similarly be used to improve the latency and throughput of chiplet networks \cite{Chiplet2018, MultiChipNoC2015} by exploring novel interconnects structures that are non-intuitive and hard for human to conceive. NoCs for domain-specific accelerators (e.g., \cite{TPU} and many others) are another possible application of the framework. Due to the data-intensive nature, accelerators can greatly benefit from high performance \cite{NoCAccel} and possibly reconfigurable \cite{WinogradAccel} NoCs, where the framework can be extended to explore better connectivity among processing elements (PEs) and between PEs and memory.

\section{Related Work}  
Research on routerless NoCs has been limited to two methods. IMR uses a genetic algorithm with random mutations to generate loop configuration. REC constructs layers recursively, generating an exact structure for a given NoC size. Our approach fundamentally differs from IMR and REC as it can guarantee fully connected loop configurations with various design constraints. This advantage is crucial to allow improved flexibility in diverse applications.

Many studies have explored machine learning applied to architecture and related tools \cite{dvfs, Ipek2006, Ipek2008, Ipek2008_2, Jimenez, NNBranchPred2011, Perceptron2016, NGPU2015, NNCMPPower, LSTMPrefetch, BayesianAccel, GPUProgOpt}, but none have explored application to routerless NoCs. Performance prediction, for example, is a popular topic for machine learning application, e.g., Ipek et al. \cite{Ipek2006, Ipek2008} pair artificial neural networks with sigmoid activation functions to build a predictive design-space model. Machine learning has also been applied to architectural components, e.g., Jim\'enez et al. \cite{Perceptron2016} use a perceptron-based approach for last level cache reuse prediction. Similar research is limited to specific aspects of architectural design and is thus complementary to our work on routerless NoCs.

Machine learning has also been used to address NoC design concerns such as congestion. Ipek et al. \cite{Ipek2008_2} use reinforcement learning to mitigate traffic congestion with an approximate return function. The learned function allowed improved path selection for packet transfer using current traffic statistics such as queue lengths. That work, however, uses a single learned function and does not enforce specific design constraints. In contrast, our framework involves both a policy and value function, using a two-headed DNN structure, both of which are subject to strict design constraints.

\section{Conclusion}
Design space exploration using deep reinforcement learning promises broad application to architectural design. Current Routerless NoC designs, in particular, have been limited by their ability to search design space, making routerless NoCs an ideal case study to demonstrate our innovative framework. The proposed framework integrates deep learning and Monte Carlo search tree with multi-threaded learning to efficiently explore large design space under constraints. Full system simulations shows that, compared with state-of-the-art routerless NoC, our proposed deep reinforcement learning NoC can achieve a 1.47x increase in throughput, 1.18X reduction in packet latency, and 1.14x reduction in average hop count, with only a few percent of power overhead. The proposed framework has broad applicability to other NoC design exploration problems with constraints, and future work can be conducted to investigate this further.

\section*{ACKNOWLEDGMENT}

This research is supported, in part, by the National Science Foundation (NSF) grants \#1566637, \#1619456, \#1619472 and \#1750047, and Software and Hardware Foundations.


\bibliographystyle{IEEEtranS}
\bibliography{refs}

\end{document}